\documentclass[twocolumn, secnumaRabic, amssymb, nobibnotes, aps, showpacs,superscriptaddress]{revtex4-1}
\usepackage{graphicx,subfigure,amsmath}%
\usepackage{subfigure}
\usepackage{color}
\usepackage{bm}
\usepackage{epstopdf}
\usepackage{amsmath,amstext}
\usepackage{graphicx}
\usepackage{booktabs}
\usepackage{multirow}
\usepackage{float}

\newcommand\myemptypage{
	\null
	\thispagestyle{empty}
	\addtocounter{page}{-1}
	\newpage
}

\begin{document}

\title{Exceptional-Point Dynamics}

\author{Yan Xing}
\affiliation{School of Physics, Zhengzhou University, Zhengzhou 450001, China}
\author{Xuedong Zhao}
\affiliation{Academy for Quantum Science and Technology, School of Electronics and Information, Zhengzhou University of Light Industry, Zhengzhou 450001, China}
\author{Hui Jing}
\email{jinghui73@foxmail.com}    
\affiliation{Key Laboratory of Low-Dimensional Quantum Structures and Quantum Control of Ministry of Education, Department of Physics and Synergetic Innovation Center for Quantum Effects and Applications, Hunan Normal University, Changsha, Hunan 410081, China}
\affiliation{Academy for Quantum Science and Technology, School of Electronics and Information, Zhengzhou University of Light Industry, Zhengzhou 450001, China}
\author{Shi-Lei Su}
\email{slsu@zzu.edu.cn}
\affiliation{School of Physics, Zhengzhou University, Zhengzhou 450001, China}
\affiliation{Institute of Quantum Materials and Physics, Henan Academy of Sciences, Zhengzhou 450046, China}
\date{\today}

\begin{abstract}
Exceptional points (EPs) play a vital role in non-Hermitian (NH) systems, driving unique dynamical phenomena and promising innovative applications. However, the NH dynamics at EPs remains obscure due to the incomplete biorthogonal eigenspaces of defective NH Hamiltonians and thus is often avoided. In this manuscript, we systematically establish pseudo-completeness relations at EPs by employing all available generalized eigenstates, where both single and multiple arbitrary-order EPs embracing degenerate scenarios are addressed, to unveil EP dynamics. We reveal that depending on EP order and initial conditions, the EP dynamics is characterized by a \emph{polynomial growth over time} of coalescing eigenstates or their superposition, which will dominate long-term evolution despite real spectra protected by pseudo-Hermiticity (PH), or can also become unitary. We further introduce two PH-compliant NH models to demonstrate these EP dynamics and explore their applications. This work completes the dynamical investigation of NH physics, offers valuable insights into the nonunitary evolution at EPs, and further lays the foundation for engineering and exploiting EP-related devices and technologies.
\end{abstract}

\maketitle

\emph{Introduction}.---In traditional quantum mechanics, Hermitian operators are essential since they guarantee the reality of eigenvalues, corresponding to observables. As a result, non-Hermitian (NH) Hamiltonians~\cite{Moiseyev2011,Bagarello2016,Ashida02072020}, which effectively mimic dissipative open systems, were historically overlooked until the seminal discovery of parity-time ($\mathcal{PT}$) symmetry~\cite{PhysRevLett.80.5243,RevModPhys.96.045002}. This symmetry enables certain NH operators to support real eigenvalues, challenging the constraint of Hermiticity. With the enhancement of NH level, these $\mathcal{PT}$-symmetric systems can undergo a spontaneous symmetry-breaking transition, engendering the reality-complexity conversion of spectrum. At the criticality, a spectral singularity, termed exceptional point (EP)~\cite{Heiss2001,Heiss2004,PhysRevA.72.014104,PhysRevA.81.022102,Heiss2010,Heiss2012,PhysRevLett.125.203602,PhysRevA.94.053834,PhysRevA.95.022125,PhysRevA.96.043821,PhysRevA.101.033820,PhysRevA.107.062209,PhysRevB.100.144301,PhysRevB.101.224301,PhysRevB.108.115427,PhysRevResearch.2.033127,PhysRevA.104.063508,Jing2017,Zhou2018,Ashida2017,PhysRevResearch.2.033018,Hashemi2022,Zhang2025,PhysRevB.111.075171,PhysRevB.111.L100305,wang2023measurement,PhysRevA.110.012226,agarwal2023recognizing}, emerges. Unlike diabolic points (DPs) for conventional degeneracy, EPs belong to nontrivial degeneracy where all corresponding eigenstates coalesce, leading to the skewness of eigenspace. This peculiarity has been leveraged in multifarious realms to yield a plethora of exotic phenomena and absorbing applications without or surpassing the Hermitian counterparts. On the one hand, in the proximity of EPs, e.g., enhanced sensitivity~\cite{Hodaei2017,Chen2017,Hokmabadi2019,Kononchuk2022,Park2020,PhysRevLett.122.153902,PhysRevLett.123.180501,PhysRevLett.123.213901,PhysRevLett.128.173602,PhysRevLett.130.227201,PhysRevLett.131.220801,PhysRevLett.132.243601,PhysRevLett.134.133801,Zhao2018,Lau2018,Wang2020,Wang2020NC,Yang2023,Suntharalingam2023,Chen2024,Zhao2024,Mao2024,18gg-gvzc}, entanglement transition and accelerated generation~\cite{PhysRevLett.131.260201,PhysRevLett.131.100202}, unconventional wave propagation~\cite{Wang2021,Wang2020NP,PhysRevLett.121.093901,PhysRevLett.123.214302,PhysRevLett.124.253202,PhysRevLett.133.173801,PhysRevX.8.031035,Soleymani2022,Zhu2024,Xu2023,Mao2025}, and controllable laser~\cite{Liao2023,Guilhem2024} have been found. On the other hand, harnessed by encircling, chiral mode switching~\cite{Doppler2016,Schumer2022,Ren2022,PhysRevLett.118.093002,PhysRevLett.124.153903,PhysRevLett.125.187403,PhysRevLett.126.170506,PhysRevLett.127.253901,PhysRevLett.128.110402,PhysRevLett.128.160401,PhysRevLett.129.127401,PhysRevLett.129.273601,PhysRevLett.132.243802,PhysRevLett.133.070403,PhysRevLett.133.053802,PhysRevLett.133.113802,PhysRevLett.133.261401,PhysRevLett.134.146602,PhysRevX.8.021066,Shu2022,Arkhipov2023}, nonreciprocal transmission~\cite{Choi2017,PhysRevLett.132.243602}, improved quantum heat engine~\cite{PhysRevLett.130.110402}, etc. have been also proposed. Additionally, pseudo-Hermiticity (PH)~\cite{PhysRevX.9.041015}, a generalization of $\mathcal{PT}$ symmetry, provides a universal characterization for the reality of spectrum, and NH Hamiltonians respecting PH can only admit real spectra under specific conditions, which is attributed to their similarity to the spectrally-equivalent Hermitian counterparts. Otherwise, their eigenvalues appear instead in complex conjugate pairs.

Despite significant progress in the research concerning EPs, fundamental questions persist. Typically, the evolution of NH systems at EPs remains indigestible, as the completeness relation of standard biorthogonal basis collapses, rooting in eigenstate coalescence. This motivates the pursuit of an unambiguous generic description of EP dynamics, even for the scenarios with degenerate multiple EPs of arbitrary orders, and further the unlocking of fresh applications in terms of its dynamical hallmarks. To witness these intriguing dynamical behaviors visually, incorporating PH to maintain a real spectrum contributes to eliminate the instability of eigenstates caused by complex eigenenergies.  

In this manuscript, under a comprehensive consideration encompassing both single and (degenerate) multiple arbitrary-order EPs, we resort to all available generalized eigenstates to tailor a succession of pseudo-completeness relations (PCRs) and to further clearly capture the NH dynamics at EPs. We analytically find that coalescing eigenstates or their superposition feature a \emph{polynomial growth behavior} dependent on EP order and initial conditions during evolution, where a spontaneous state conversion to them can be observed in long-time limit when PH-induced real spectra are present. We also illustrate the EP dynamics via two NH models respecting PH, highlighting the potential applications of the spontaneous conversion process and advancing the prospects of the EP dynamics in NH systems with PH.

\emph{Conventional NH dynamics invalid for EPs}.---We consider a general time-independent NH Hamiltonian $\hat{H}$ of dimension $M\times M$. The wave function evolves as $|\Psi(t)\rangle=e^{-i\hat{H}t}|\Psi(0)\rangle$ ($\hbar=1$). For a nondefective $\hat{H}$, its left and right eigenstates $\langle L_m|$ and $|R_m\rangle$ with eigenenergies $E_m$ together form a biorthogonal basis, satisfying the orthonormalization condition $\langle L_m|R_{m'}\rangle/(\sqrt{\langle L_m|R_m\rangle}\sqrt{\langle L_{m'}|R_{m'}\rangle})=\delta_{m,m'}$ ($m,m^{\prime}\in[1,M]$) and the completeness relation $\sum_{m=1}^M|R_m\rangle\langle L_m|/\langle L_m|R_m\rangle=\hat{\mathbb{I}}$, where $\delta$ and $\hat{\mathbb{I}}$ are the Kronecker symbol and the identity matrix, further allowing the straightforward analysis of NH dynamics.

However, the framework breaks down at an EP of order $N$, where $\hat{H}$ becomes defective. To see this clearly, consider the Jordan decomposition $\hat{S}_\bot^{-1}\hat{H}\hat{S}_\bot=\hat{H}_J$. Here, the Jordan canonical form $\hat{H}_J=\hat{J}\oplus\hat{D}$, the elements of the Jordan block $\hat{J}$ are  $\hat{J}_{n,n'}=E_{\mathrm{EP}}\delta_{n,n'}+\delta_{n,n'-1}$ ($n,n'\in[1,N]$) and the diagonal matrix $\hat{D}$ has elements $\hat{D}_{j,j'}=E_j\delta_{j,j'}$ ($j,j'\in [1,M-N]$), with $E_{\mathrm{EP}}$ and $E_{j}$ (first assumed nondegenerate and distinct from $E_{\mathrm{EP}}$) the coalescing eigenenergy and the remaining eigenenergies. The similarity transformation matrix $\hat{S}_\bot=(|\psi_1\rangle,\dots,|\psi_N\rangle,|\phi_1\rangle,\dots,|\phi_{M-N}\rangle)$ consists of the eigenstate $|\psi_1\rangle$ of $E_{\mathrm{EP}}$, the generalized eigenstates $|\psi_{n_1}\rangle$ ($n_1\in[2,N]$) determined by the Jordan chain $(\hat{H}-E_{\mathrm{EP}})|\psi_{n_1}\rangle=|\psi_{n_1-1}\rangle$, and the eigenstates $|\phi_j\rangle$ of $E_j$. While $|\phi_j\rangle$ and their left partners $\langle\tilde{\phi}_j|$ retain the standard orthonormalization condition and $\langle\tilde{\phi}_{j}|\psi_{1}\rangle=\langle\tilde{\psi}_{1}|\phi_{j}\rangle=0$, the self-orthogonality $\langle\tilde{\psi}_1|\psi_1 \rangle=0$ renders the inverse of the biorthogonal norm divergent, invalidating the standard completeness relation and thus impeding the analytical exploration of the NH dynamics at EPs.

\emph{EP dynamics}.---To tackle the problem, it is necessary to reestablish an identity matrix based on these currently available eigenstates and generalized eigenstates, if possible. We now introduce the matrix $\hat{\mathbb{H}}=(\hat{H}-E_{\mathrm{EP}})^N$ that is diagonalizable by $\hat{S}_\bot$ and can examine $\hat{\mathbb{H}}|\psi_n\rangle=0$ and $\hat{\mathbb{H}}|\phi_j\rangle=(E_j-E_{\mathrm{EP}})^N|\phi_j\rangle$, with similar properties for $\langle\tilde{\psi}_n|$ and $\langle\tilde{\phi}_j|$, which implies that $\{\langle\tilde{\psi}_{n}|,|\psi_{n}\rangle,\langle\tilde{\phi}_{j}|,|\phi_{j}\rangle\}$ can actually together form a biorthogonal basis in the context of $\hat{\mathbb{H}}$ with nondefective degeneracy. Accordingly, for the eigenstates $|\psi_n\rangle$ of the $N$-fold degenerate zero eigenenergy, we need to seek out their left partners, whereas the left partner of each $|\phi_{j}\rangle$ remains unchanged. Given the known biorthogonality $\langle\tilde{\psi}_{N+1-l}|\psi_{n_{1}-1}\rangle=0$ ($l\in[n_{1},N]$) and the unknown $\langle\tilde{\psi}_{N+1-p}|\psi_{n}\rangle$ ($p\in[1,n]$), the matching left partner of $|\psi_{n}\rangle$ must be $\langle\tilde{\psi}_{N+1-n}|$ and the overlaps $\langle\tilde{\psi}_{N+1-q}|\psi_{n_{1}}\rangle$ ($q\in[1,n_{1}-1]$) are required to be vanishing simultaneously.

Ultimately, we can redefine a set of left and right eigenstates subjected to the orthonormalization condition as $\{\langle\tilde{\nu}_{N+1-n}|,|\nu_{n}\rangle,\langle\tilde{\omega}_{j}|,|\omega_{j}\rangle\}$; See Supplemental Material (SM) for their concrete expressions~\cite{SM}, where $\langle\tilde{\nu}_{n}|\nu_{n^{\prime}}\rangle=\delta_{n,N+1-n^{\prime}}$, $\langle\tilde{\omega}_{j}|\omega_{j^{\prime}}\rangle=\delta_{j,j^{\prime}}$, and $\langle\tilde{\omega}_{j}|\nu_{n}\rangle=\langle\tilde{\nu}_{n}|\omega_{j}\rangle=0$, further enabling the following modified closure relation 
\begin{equation}\label{Eq-1}
\sum_{n=1}^{N}|\nu_{n}\rangle\langle\tilde{\nu}_{N+1-n}|+\sum_{j=1}^{M-N}|\omega_{j}\rangle\langle\tilde{\omega}_{j}|=\hat{\mathbb{I}},
\end{equation}
which we deem as the PCR in the context of $\hat{H}$. Note that $|\tilde{\nu}_{n}\rangle$ and $|\nu_{n}\rangle$ still obey the Jordan chain; See SM for more details of the scenario with a single EP~\cite{SM}.

With the aid of the PCR, the NH dynamics at the $N$th-order EP can be analytically evaluated as $|\Psi(t)\rangle=e^{-i\hat{H}t}(\sum_{n=1}^{N}\zeta_{n}|\nu_{n}\rangle+\sum_{j=1}^{M-N}\iota_{j}|\omega_{j}\rangle)$, where the prefactors $\zeta_{n}=\langle\tilde{\nu}_{N+1-n}|\Psi(0)\rangle$ and $\iota_{j}=\langle\tilde{\omega}_{j}|\Psi(0)\rangle$. By employing $e^{-i\hat{H}t}|\nu_{n}\rangle=e^{-iE_{\mathrm{EP}}t}\sum_{p=1}^{n}\frac{(-it)^{n-p}}{(n-p)!}|\nu_{p}\rangle$, the resulting $|\Psi(t)\rangle$ reads 
\begin{equation}\label{Eq-2}
\begin{split}
|\Psi(t)\rangle=\sum_{n=1}^{N}\sum_{p^{\prime}=0}^{N-n}\mathcal{C}_{n,p^{\prime}}\left(t\right)t^{p^{\prime}}|\nu_{n}\rangle+\sum_{j=1}^{M-N}\mathcal{C}_{j}\left(t\right)|\omega_{j}\rangle,
\end{split}
\end{equation}
with $\mathcal{C}_{n,p^{\prime}}(t)=(-i)^{p^{\prime}}\zeta_{n+p^{\prime}}e^{-iE_{\mathrm{EP}}t}/p^{\prime}!$ and $\mathcal{C}_{j}\left(t\right)=\iota_{j}e^{-iE_{j}t}$.

In the $N$th-order EP dynamics governed by Eq.~(\ref{Eq-2}), each $|\nu_n\rangle$ is equipped with a \emph{polynomial growth over time}, whose degree can be tuned by switching off $\zeta_{n+p^{\prime}}$ in descending order that depends on the selection of $|\Psi(0)\rangle$, and $|\nu_1\rangle$ always supports the highest degree. Consequently, if the NH system with PH admits a real spectrum at the EP, $|\nu_1\rangle$ will grow most rapidly and become dominant gradually during evolution as long as one of $\zeta_{n_1}\neq0$, whereas the rest of states will either grow more slowly or remain bounded. In the long-time limit $t\to\infty$, $|\Psi(t)\rangle$ will only contain the superposition of $|\nu_1\rangle$, triggering a spontaneous state conversion. Additionally, the evolution will remain unitary once all $\zeta_{n_1}=0$.

We further extend the single $N$th-order EP dynamics, as described in Eq.~(\ref{Eq-2}), to that at multiple arbitrary-order EPs, including diverse degenerate scenarios. For each scenario, we also tailor the matching PCR to analytically solve the EP dynamics, with detailed derivations available in SM~\cite{SM}. 

For multiple EPs without degeneracy, we redefine a set of orthonormalized $\{\langle\tilde{\nu}_{N_s+1-y_s}^{(s)}|, |\nu_{y_s}^{(s)}\rangle, \langle\tilde{\omega}_r|, |\omega_r\rangle\}$, where $s\in[1,\mathbb{N}]$ indexes these EPs with $\mathbb{N}$ counting the number of them, $y_s\in[1,N_s]$ with $N_s$ giving the order of the $s$th EP, and $r\in[1,\mathbb{M}]$ with $\mathbb{M}=M-\sum_{s=1}^{\mathbb{N}}N_s$, meeting $\langle \tilde{\nu}_{y_s}^{(s)}|\nu_{y_{s'}'}^{(s')}\rangle=\delta_{s,s'}\delta_{y_s,N_s+1-y_{s'}'}$, $\langle\tilde{\omega}_r|\omega_{r'}\rangle=\delta_{r,r'}$, and $\langle \tilde{\omega}_r|\nu_{y_s}^{(s)}\rangle=\langle\tilde{\nu}_{y_s}^{(s)}|\omega_r\rangle=0$. The PCR now reads
\begin{equation}\label{Eq-1-1}
\begin{split}
\sum_{s=1}^{\mathbb{N}}\sum_{y_{s}=1}^{N_{s}}|\nu_{y_{s}}^{(s)}\rangle\langle\tilde{\nu}_{N_{s}+1-y_{s}}^{(s)}|+\sum_{r=1}^{\mathbb{M}}|\omega_{r}\rangle\langle\tilde{\omega}_{r}|=\hat{\mathbb{I}},
\end{split}
\end{equation} 
simplifying to Eq.~(\ref{Eq-1}) if $\mathbb{N}=1$. We can thus obtain
\begin{equation}\label{Eq-1-2}
|\Psi(t)\rangle=\sum_{s=1}^{\mathbb{N}}\sum_{y_{s}=1}^{N_{s}}\sum_{p_{s}^{\prime}=0}^{N_{s}-y_{s}}\mathcal{C}_{y_{s},p_{s}^{\prime}}^{(s)}\left(t\right)t^{p_{s}^{\prime}}|\nu_{y_{s}}^{(s)}\rangle+\sum_{r=1}^{\mathbb{M}}\mathcal{C}_{r}\left(t\right)|\omega_{r}\rangle,
\end{equation}
where $\mathcal{C}_{y_s,p_s'}^{(s)}(t)=(-i)^{p_s'}\zeta_{y_s+p_s'}^{(s)}e^{-i E_{\mathrm{EP}}^{(s)}t}/p_s'!$ with $\zeta_{y_s}^{(s)}=\langle\tilde{\nu}_{N_s+1-y_s}^{(s)}|\Psi(0)\rangle$, and $\mathcal{C}_r(t)=\iota_re^{-i E_r t}$ with $\iota_r=\langle\tilde{\omega}_r|\Psi(0)\rangle$. See SM for the scenarios of multiple EPs with degeneracy and additional nondefective degeneracy~\cite{SM}.

It turns out that in PH systems presenting real spectra at multiple EPs, the long-time dynamics is governed by the coalescing eigenstates with the highest polynomial growth. Specifically, as long as the prefactor of $t^{\max(N_s-1)}$ is nonvanishing, $|\Psi(t)\rangle$ converges to the eigenstate of the highest-order EP or a superposition if there are some EPs with the highest order, where the superposition coefficients are determined by the relevant prefactors and eigenenergies. Otherwise, the dynamics for $t\to\infty$ is instead dominated by the next-highest growth terms. When all polynomial terms vanish (keeping at most the prefactors of $t^0$), the evolution reduces to unitary dynamics.

These findings suggest a prospect of engineering passive quantum devices based on the spontaneous state conversion process and provide a flexible route to manipulate the steady state in a system with multiple EPs by judiciously choosing the initial conditions. To clarify this, we exemplify two NH models with PH, authenticating the predicted dynamical behavior and emphasizing their advantages over conventional schemes.

\begin{figure}
	\centering
	\includegraphics[width=1.0\linewidth]{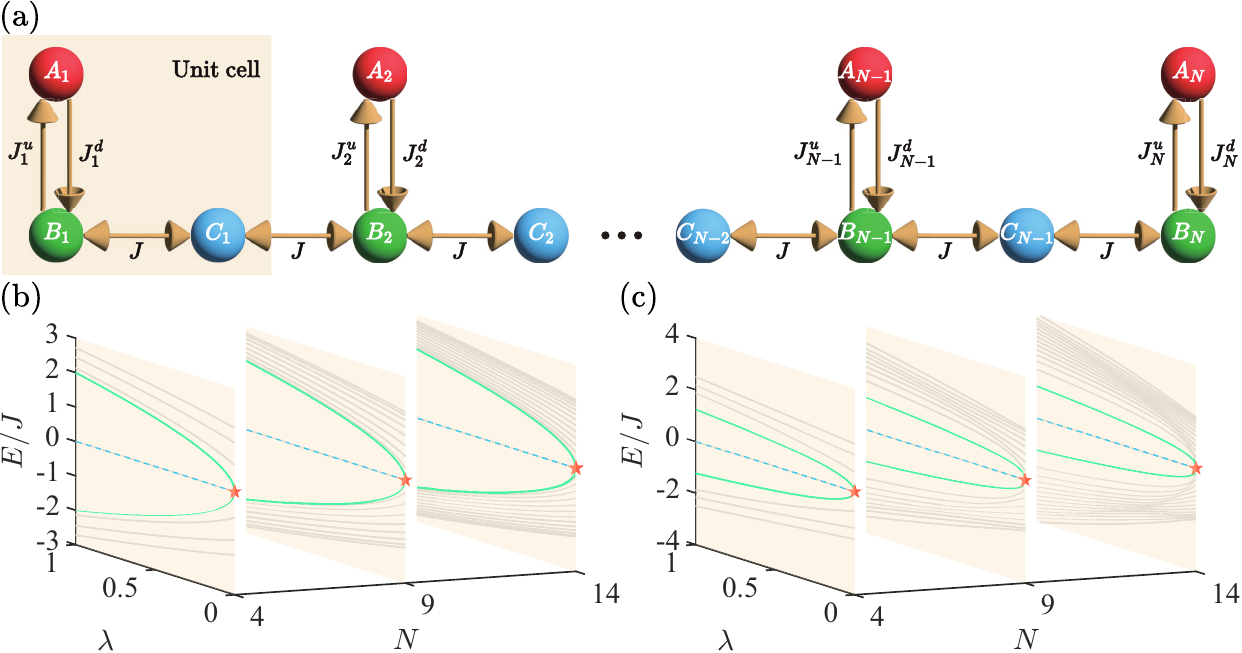}
	\caption{(a) Schematic of NH stub ribbon with PH. (b) Spectra with different $N$ for $\{J_n^u=2J\}$ as a function of $\lambda$. Gray-solid and blue-dashed lines represent the real and imaginary parts, where the zero eigenenergy is $(N-1)$-fold degenerate for $\lambda\in(0,1)$. The green-solid line delineates the pair of eigenenergies with the minimum absolute value, merging into zero energy at $\lambda=0$ to form a 2nd-order EP. The red marker labels the hybridization of the $(N-1)$-order DP and the 2nd-order EP at $\lambda=0$. (c) Same as (b) but for $\{J_n^u=\sqrt{n}J\}$.}
	\label{fig1}
\end{figure}

\emph{Spontaneous arbitrary-ratio multiport beam splitter}.---We first investigate a NH ramification of the stub ribbon~\cite{Flach_2014,PhysRevA.92.052103,PhysRevB.96.064305,Ge:18,PhysRevResearch.6.L022006}, a prototypical quasi-one-dimensional flat-band model, as depicted in Fig.~\ref{fig1}(a). Each unit cell is made up of three sublattices, $A$, $B$, and $C$, except for the last unit cell in the absence of $C$. $N\geq2$ denotes the number of unit cells. The Hamiltonian reads 
\begin{equation}\label{Eq-3}
\hat{H}=\sum_{n}J_{n}^{u}\hat{a}_{n}^{\dagger}\hat{b}_{n}+J_{n}^{d}\hat{b}_{n}^{\dagger}\hat{a}_{n}+J\left(\hat{c}_{n}^{\dagger}\hat{b}_{n}+\hat{c}_{n}^{\dagger}\hat{b}_{n+1}+\mathrm{H.c.}\right),
\end{equation}
where $\hat{a}_n^\dagger$, $\hat{b}_n^\dagger$, and $\hat{c}_n^\dagger$ are creation operators for a particle on sites $A_n$, $B_n$, and $C_n$ in the $n$th unit cell. While both the intracell and intercell hoppings between $B$ and $C$ are reciprocal with uniform strength $J$, the upward and downward hoppings between $A$ and $B$ inside the same unit cell $n$ are nonreciprocal with strengths $J_n^u$ and $J_n^d$, endowing the system with non-Hermiticity. When $J_n^u=J_n^d$ for all $n\in[1,N]$, the model reduces to its Hermitian counterpart. A concise introduction of $(N-1)$-fold degenerate zero-energy flat band and associated compact localized states for the Hermitian situation is provided in SM~\cite{SM}.

In the NH regime of $J_n^u\neq J_n^d$, we assign, without loss of generality, $J_n^u>0$ and $J_n^d =\lambda J_n^u$ with $\lambda\in[0,1)$. The Hermitian situation corresponds to $\lambda=1$. For $\lambda\in(0,1)$, the system continues to admit a real spectrum, inherently stemming from the respect of PH, $\hat{\eta}^{-1}\hat{H}\hat{\eta}=\hat{H}^\dagger$, where
\begin{equation}\label{Eq-4}
\begin{split}
\hat{\eta}=\left[\bigoplus_{j=1}^{N-1}\operatorname{diag}\left(\dfrac{J_j^u}{J_j^d},1,1\right)\right]\bigoplus\operatorname{diag}\left(\dfrac{J_N^u}{J_N^d},1\right).
\end{split}
\end{equation}
The reality of the spectrum also reflects the spectral equivalence of the model to a Hermitian stub ribbon governed by $\hat{H}_h=\sum_{n}\sqrt{J_n^u J_n^d}(\hat{a}_n^\dagger\hat{b}_n +\hat{b}_n^\dagger\hat{a}_n)+J(\hat{c}_n^\dagger\hat{b}_n+\hat{b}_n^\dagger\hat{c}_n+\hat{c}_n^\dagger\hat{b}_{n+1}+\hat{b}_{n+1}^\dagger \hat{c}_n)$, via the similarity transformation $\hat{S}^{-1}\hat{H}\hat{S}=\hat{H}_h$, where $\hat{S}=\sqrt{\hat{\eta}}$. Chiral symmetry$^{\dagger}$ in the NH framework~\cite{PhysRevX.9.041015}, $\hat{\Gamma}\hat{H}\hat{\Gamma}^{-1}=-\hat{H}$, with
\begin{equation}\label{Eq-5}
\begin{split}
\hat{\Gamma}=\left[\bigoplus_{j=1}^{N-1}\operatorname{diag}\left(1,-1,1\right)\right]\bigoplus\operatorname{diag}\left(1,-1\right),
\end{split}
\end{equation}
further produces a symmetric real spectrum. The $(N-1)$-fold degenerate zero-energy flat band persists, with right eigenstates $|\xi_l^R\rangle=\frac{1}{J_l^d}|A_l\rangle+\frac{1}{J_{l+1}^d}|A_{l+1}\rangle-\frac{1}{J}|C_l\rangle$ ($l\in[1,N-1]$), corresponding to NH compact localized states.

As $\lambda$ diminishes, for these upper and lower bulk bands, the absolute values of their eigenenergies dwindle and once $\lambda=0$, a pair of eigenenergies with the minimum absolute value among them merges into zero energy, forming a 2nd-order EP and further rendering the spectrum singular. Therefore, a hybridization of the $(N-1)$-order DP and the 2nd-order EP arises. Figures~\ref{fig1}(b) and~\ref{fig1}(c) show the spectra with different $N$ versus $\lambda$ for $\{J_n^u=2J\}$ and $\{J_n^u=\sqrt{n}J\}$. Visibly, the $(N+1)$-fold degeneracy of the zero eigenenergy at $\lambda=0$ is independent of $N$ and $\{J_n^u\}$. See SM for these spectra with $N=4$ in a broader scope of $\lambda$, where three 3rd-order EPs occur successively when $\lambda<0$, and the identification of these EPs via the Petermann factor~\cite{SM}.

At the 2nd-order EP, the $N-1$ zero-energy eigenstates become extremely localized due to $J_n^d=0$ and take the form $|\xi_l^{\mathrm{EP}}\rangle=|A_{l+1}\rangle$, with left eigenstates $\langle\tilde{\xi}_l^{\mathrm{EP}}|=\frac{1}{J_l^u}\langle A_l|+\frac{1}{J_{l+1}^u}\langle A_{l+1}|-\frac{1}{J}\langle C_l|$. The extra coalescing zero-energy eigenstate and the associated generalized eigenstate can be derived as $|\chi^{\mathrm{EP}}\rangle=\sum_{n=1}^{N}(-1)^{n+1}J_n^u|A_n\rangle$ and $|\chi_a^{\mathrm{EP}}\rangle=\sum_{n=1}^{N}(-1)^{n+1}|B_n\rangle$, with their left partners $\langle\tilde{\chi}^{\mathrm{EP}}|=\sum_{n=1}^{N}(-1)^{n+1}\langle B_n|$ and $\langle\tilde{\chi}_{a}^{\mathrm{EP}}|=\frac{N}{J_{1}^{u}}\langle A_{1}|+\sum_{n=1}^{N-1}(-1)^{n}\frac{(N-n)}{J}\langle C_{n}|$. 

We can now find that the degeneracy disrupts the biorthogonality $\langle\tilde{\xi}_l^{\mathrm{EP}}|\xi_{l'}^{\mathrm{EP}}\rangle=0$ for $l\neq l'$. According to previous universal theory, we thus rewrite $\langle\tilde{\xi}_l^{\mathrm{EP}}|$ as $\langle\tilde{\varrho}_l^{\mathrm{EP}}|=\sum_{s=1}^{l}(-1)^{l+s}\langle\tilde{\xi}_s^{\mathrm{EP}}|$ and further redefine $|\nu^{\mathrm{EP}}\rangle=|\chi^{\mathrm{EP}}\rangle/\sqrt{N}$, $|\nu_{a}^{\mathrm{EP}}\rangle=|\chi_{a}^{\mathrm{EP}}\rangle/\sqrt{N}$, $|\mu_{l}^{\mathrm{EP}}\rangle=\sqrt{J_{l+1}^{u}}|\xi_{l}^{\mathrm{EP}}\rangle$, $|\omega_{f}^{\mathrm{EP}}\rangle=|\phi_{f}^{\mathrm{EP}}\rangle/\sqrt{\langle\tilde{\phi}_{f}^{\mathrm{EP}}|\phi_{f}^{\mathrm{EP}}\rangle}$, and their left partners, where $|\phi_{f}^{\mathrm{EP}}\rangle$ ($f\in[1,2N-2]$) are the remaining bulk states, to establish the PCR,
\begin{equation}\label{Eq-6}
\begin{split}
&|\nu^{\mathrm{EP}}\rangle\langle\tilde{\nu}_{a}^{\mathrm{EP}}|+|\nu_{a}^{\mathrm{EP}}\rangle\langle\tilde{\nu}^{\mathrm{EP}}|\\
&+\sum_{l=1}^{N-1}|\mu_{l}^{\mathrm{EP}}\rangle\langle\tilde{\mu}_{l}^{\mathrm{EP}}|+\sum_{f=1}^{2N-2}|\omega_{f}^{\mathrm{EP}}\rangle\langle\tilde{\omega}_{f}^{\mathrm{EP}}|=\hat{\mathbb{I}}.
\end{split}
\end{equation} 
Eventually, the 2nd-order EP dynamics can be expressed as 
\begin{equation}\label{Eq-7}
\begin{split}
|\Psi\left(t\right)\rangle=&\left(\zeta^{\mathrm{EP}}-it\zeta_{a}^{\mathrm{EP}}\right)|\nu^{\mathrm{EP}}\rangle+\zeta_{a}^{\mathrm{EP}}|\nu_{a}^{\mathrm{EP}}\rangle\\
&+\sum_{l=1}^{N-1}\tau_{l}^{\mathrm{EP}}|\mu_{l}^{\mathrm{EP}}\rangle+\sum_{f=1}^{2N-2}\iota_{f}^{\mathrm{EP}}e^{-iE_{f}t}|\omega_{f}^{\mathrm{EP}}\rangle,
\end{split}
\end{equation} 
with $\zeta^{\mathrm{EP}}=\langle \tilde{\nu}_a^{\mathrm{EP}}|\Psi(0)\rangle$, $\zeta_a^{\mathrm{EP}}=\langle\tilde{\nu}^{\mathrm{EP}}|\Psi(0)\rangle$, $\tau_{l}^{\mathrm{EP}}=\langle\tilde{\mu}_{l}^{\mathrm{EP}}|\Psi(0)\rangle$, and $\iota_{f}^{\mathrm{EP}}=\langle\tilde{\omega}_{f}^{\mathrm{EP}}|\Psi(0)\rangle$.

When $\zeta_a^{\mathrm{EP}}\neq0$, requiring an initial nonvanishing occupation at least on one of the sublattices $B$, $|\nu^{\mathrm{EP}}\rangle$ will grow linearly over time, automatically distributing population only across all sublattices $A$ in the ratio $(J_1^u)^2:\cdots:(J_N^u)^2$. If we treat an arbitrary site $B_n$ that one excitation is initially injected into as an input port and all sites $A_n$ as $N$ output ports, the NH stub ribbon at the 2nd-order EP can serve as a spontaneous beam splitter, where the splitting ratio is tunable via $\{J_n^u\}$ and the number of output ports is dictated by $N$, in contrast to the schemes of the directional adiabatic modulation of parameters within a specified range~\cite{PhysRevB.103.085129,PhysRevB.109.094303} and of two or several ports with only equal ratios~\cite{PhysRevLett.126.230503,Wang2024Nature}.

\begin{figure}
	\centering
	\includegraphics[width=1.0\linewidth]{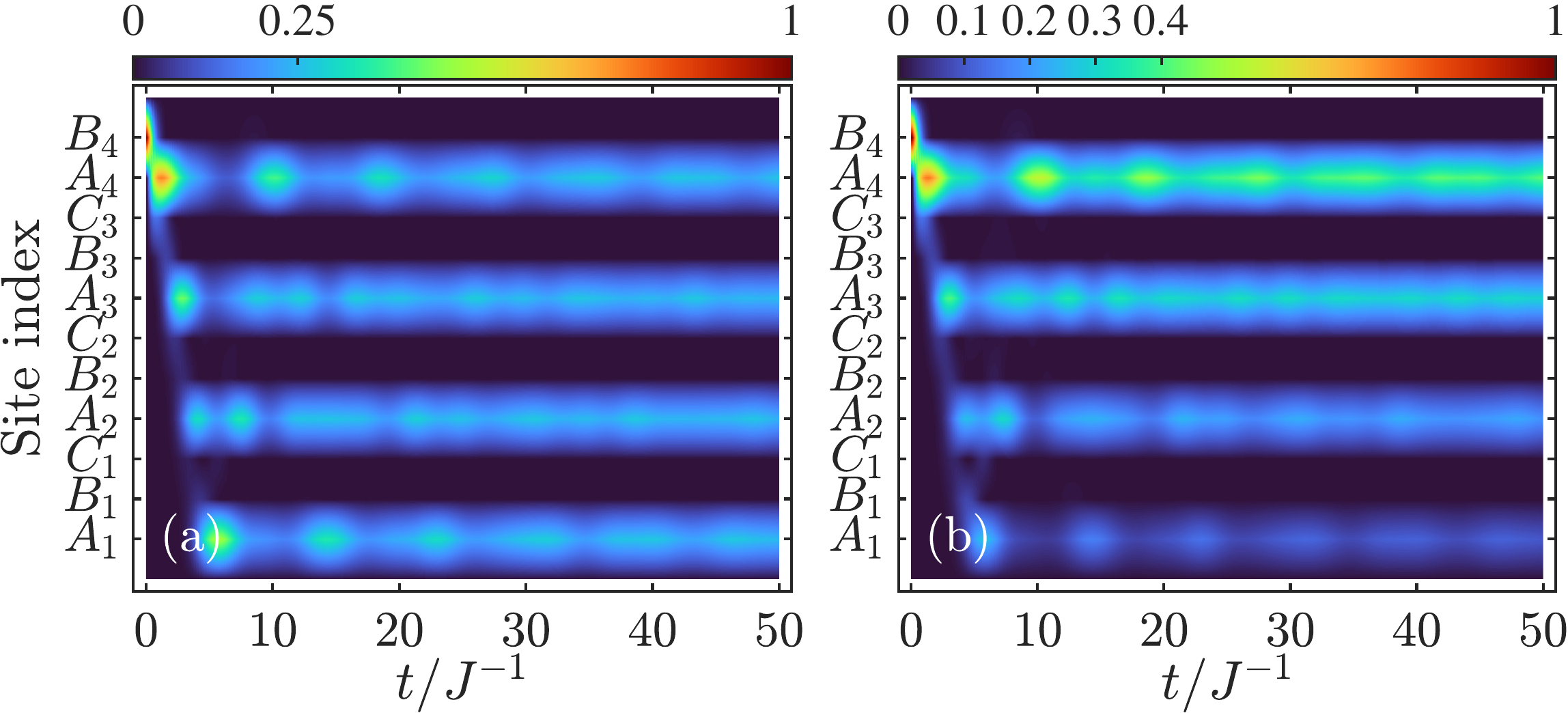}
	\caption{(a) Population dynamics of normalized $|\Psi(t)\rangle$ with $N=4$, $\lambda=0$, $\{J_n^u=2J\}$, and $|\Psi(0)\rangle=|B_4\rangle$, showing a 4-port spontaneous splitting process with ratio $25:25:25:25$. (b) Same as (a) but for $\{J_n^u=\sqrt{n}J\}$, yielding a ratio of $10:20:30:40$.}
	\label{fig2}
\end{figure}

For $N=4$ and $\lambda=0$, figure~\ref{fig2}(a) shows the 2nd-order EP dynamics when $\{J_n^u=2J\}$ and $|\Psi(0)\rangle=|B_4\rangle$, achieving a 4-port spontaneous beam splitter with an equal ratio of $25:25:25:25$. Figure~\ref{fig2}(b) is the same as Fig.~\ref{fig2}(a) except for $\{J_n^u=\sqrt{n}J\}$ and an arithmetic splitting ratio of $10:20:30:40$ can be attained. The marginal oscillations aroused by bulk states become negligible as $t\to\infty$. See SM for the spontaneous splitting dynamics of the two ratios with different initial states in longer-term evolution and the spontaneous conversion process to $|\nu^{\mathrm{EP}}\rangle$ corroborated by fidelity~\cite{SM}.

\begin{figure*}
	\centering
	\includegraphics[width=1.0\linewidth]{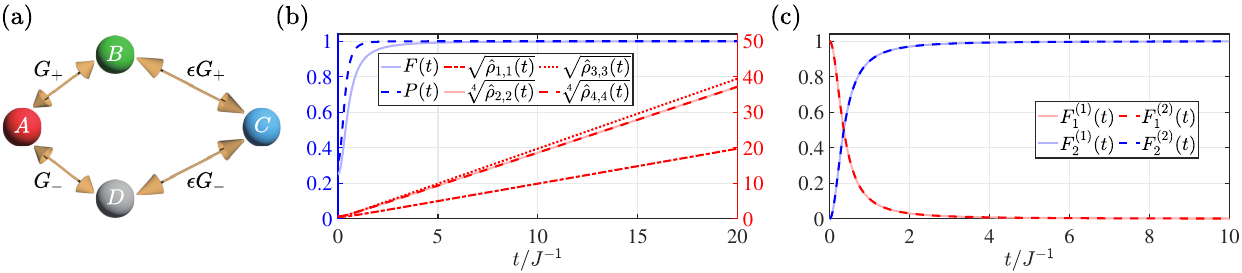}
	\caption{(a) Schematic of NH diamond ring with $\mathcal{PT}$ symmetry. (b) Spontaneous two-body entanglement generation from a completely mixed state $\hat{\rho}(0)$ for $\epsilon=2$ and $\kappa=1$, showing average fidelity $F(t)$, purity $P(t)$, and diagonal elements of $\hat{\rho}(t)$ over 1000 random initial states. (c) Spontaneous two-body entanglement transfer manifested by fidelities $F_{1}^{(1)}(t)$, $F_{2}^{(1)}(t)$ ($\epsilon\neq\pm i$, $\kappa=1$ to $\epsilon=-i$, $\kappa=-1$) and $F_{1}^{(2)}(t)$, $F_{2}^{(2)}(t)$ ($\epsilon=i$, $\kappa\neq\pm1$ to $\epsilon=-i$, $\kappa=-1$).}\label{fig3}
\end{figure*}

\emph{Spontaneous entanglement generation and transfer}.---We next turn to the second NH model characterized by a diamond-shaped ring configuration with complex coupling, as depicted in Fig.~\ref{fig3}(a). The system is governed by the Hamiltonian 
\begin{equation}\label{Eq-8}
\begin{split}
\hat{H}=&G_{+}\left(\hat{a}^{\dagger}\hat{b}+\hat{b}^{\dagger}\hat{a}\right)+\epsilon G_{+}\left(\hat{b}^{\dagger}\hat{c}+\hat{c}^{\dagger}\hat{b}\right)\\
&+\epsilon G_{-}\left(\hat{c}^{\dagger}\hat{d}+\hat{d}^{\dagger}\hat{c}\right)+G_{-}\left(\hat{d}^{\dagger}\hat{a}+\hat{a}^{\dagger}\hat{d}\right).
\end{split}
\end{equation}
Here, $\hat{a}$, $\hat{b}$, $\hat{c}$, and $\hat{d}$ are annihilation operators for modes $A$, $B$, $C$, and $D$. The complex coupling strengths $G_{\pm}=J\pm i\kappa$ with real $\kappa$ the gain/loss rate and $\epsilon$ is a real or imaginary tunable parameter. We refer to $J=1$ as the energy unit. The model respects $\mathcal{PT}$ symmetry, $[\mathcal{\hat{P}\hat{T}},\hat{H}]=0$, where
\begin{equation}\label{Eq-9}
\mathcal{\hat{P}} = \operatorname{diag}\left(1,1,\vartheta,1\right)\cdot \left(1\oplus\operatorname{antidiag}\left(1,1,1\right)\right),
\end{equation}
with $\vartheta=1$ ($-1$) for real (imaginary) $\epsilon$, and $\mathcal{\hat{T}}$ takes complex conjugation operation.

The eigenenergies of the system are $E_{1,2}=\pm i\sqrt{2(1+\epsilon^2)(\kappa^2-1)}$ and $E_{3,4}=0$. For real $\epsilon$ or imaginary $\epsilon=i\epsilon'$ with $-1<\epsilon'<1$, the spectrum is real when $-1<\kappa<1$, and $E_{1,2}$ become conjugate imaginary once $\kappa<-1$ or $\kappa>1$. The case reverses if $\epsilon'<-1$ or $\epsilon'>1$. At $\kappa=\pm1$ ($\epsilon=\pm i$), all eigenenergies vanish, forming a 3rd-order EP with an extra degenerate zero eigenenergy as long as $\epsilon\neq\pm i$ ($\kappa\neq\pm1$); Otherwise, two degenerate 2nd-order EPs emerge instead. See also SM for the identification of these EPs via the Petermann factor, the redefined eigenstates and generalized eigenstates at different EPs, and the matching PCRs~\cite{SM}.

At the 3rd-order EP ($\kappa=\pm 1$, $\epsilon\neq\pm i$, or $\epsilon=\pm i$, $\kappa\neq\pm1$), the relevant dynamics spontaneously generate two-body entanglement, converging to the entangled coalescing eigenstate $|\nu_1\rangle$, which can be realized even from a mixed initial state and also does not have to be accomplished only at a given moment compared with Ref.~\cite{PhysRevLett.131.100202}. Consider a completely mixed state $\hat{\rho}(0)=\sum_{X=A,B,C,D}\mathbb{C}_X|X\rangle\langle X|$ with $\sum_{X=A,B,C,D}\mathbb{C}_X=1$. The normalized density matrix $\hat{\rho}_{\text{nor}}(t)=\hat{\rho}(t)/\operatorname{tr}(\hat{\rho}(t))$ with $\hat{\rho}(t)=e^{-i\hat{H}t}\hat{\rho}(0)e^{i\hat{H}^{\dagger}t}$ evolves toward $|\nu_1\rangle\langle\nu_1|$ as $t\to\infty$, which is measured by the fidelity $F(t)=\langle\nu_1'|\hat{\rho}_{\text{nor}}(t)|\nu_1'\rangle$ with $|\nu_1'\rangle=|\nu_1\rangle/\sqrt{\langle\nu_1|\nu_1\rangle}$ and can be also interpreted as an entanglement storage since the steady state is always $|\nu_{1}\rangle$. As an example, for real $\epsilon$ and $\kappa=1$, $F(t)=\frac{(\mathbb{C}_{B}+\mathbb{C}_{D})+8(\mathbb{C}_{A}+\mathbb{C}_{C}\epsilon^{2})t^{2}+4(\mathbb{C}_{B}+\mathbb{C}_{D})(1+\epsilon^{2})^{2}t^{4}}{2+4[2(\mathbb{C}_{A}+\mathbb{C}_{C}\epsilon^{2})+(\mathbb{C}_{B}+\mathbb{C}_{D})(1+\epsilon^{2})]t^{2}+4(\mathbb{C}_{B}+\mathbb{C}_{D})(1+\epsilon^{2})^{2}t^{4}}$, leading to $F(t)\to1$ when $t\to\infty$. The spontaneous entanglement generation can be well understood from the distributions of $|\tilde{\nu}_{1}\rangle$ and $|\tilde{\nu}_{2}\rangle$, together covering $|A\rangle$, $|B\rangle$, $|C\rangle$, and $|D\rangle$ and thus rendering the prefactors $\zeta_{3}$ or $\zeta_{2}$ always nonvanishing no matter what the initial state (not an eigenstate) is. Hence, $|\nu_{1}\rangle$ will quadratically or linearly grow over time. Figure~\ref{fig3}(b) shows $F(t)$ and the purity $P(t)=\operatorname{tr}(\hat{\rho}_{\text{nor}}^2(t))$, averaged over 1000 random initial states and further highlighting the spontaneous purification process. For clarity, we also plot the average diagonal elements $\sqrt{\hat{\rho}_{1,1}(t)}$, $\sqrt[4]{\hat{\rho}_{2,2}(t)}$, $\sqrt{\hat{\rho}_{3,3}(t)}$, and $\sqrt[4]{\hat{\rho}_{4,4}(t)}$. All the linear behaviors indicate the quadratical growth of $|\nu_{1}\rangle$ and the linear growth of $|\nu_{2}\rangle$, which can be authenticated more concretely by the same slope of $\sqrt[4]{\hat{\rho}_{2,2}(t)}$ and $\sqrt[4]{\hat{\rho}_{4,4}(t)}$ and the ratio $\epsilon$ of the slope of $\sqrt{\hat{\rho}_{3,3}(t)}$ to that of $\sqrt{\hat{\rho}_{1,1}(t)}$.

For $\epsilon=\pm i$ and $\kappa=\pm 1$, the dynamics at the two degenerate 2nd-order EPs enables a spontaneous two-body entanglement transfer, where $|\Psi(t)\rangle$ evolves as
\begin{equation}\label{Eq-12}
\begin{split}
|\Psi(t)\rangle=&(\zeta_{1}^{(1)}-it\zeta_{2}^{(1)})|\nu_{1}^{(1)}\rangle+\zeta_{2}^{(1)}|\nu_{2}^{(1)}\rangle\\
&+(\zeta_{1}^{(2)}-it\zeta_{2}^{(2)})|\nu_{1}^{(2)}\rangle+\zeta_{2}^{(2)}|\nu_{2}^{(2)}\rangle,
\end{split}
\end{equation} 
with the prefactors $\zeta_{1}^{(1)}=\langle\tilde{\nu}_{2}^{(2)}|\Psi(0)\rangle$, $\zeta_{2}^{(1)}=\langle\tilde{\nu}_{1}^{(2)}|\Psi(0)\rangle$, $\zeta_{1}^{(2)}=\langle\tilde{\nu}_{2}^{(1)}|\Psi(0)\rangle$, and $\zeta_{2}^{(2)}=\langle\tilde{\nu}_{1}^{(1)}|\Psi(0)\rangle$. As an example, for $\epsilon=-i$, $\kappa=-1$, and an initial entangled state $|\Psi(0)\rangle=\frac{1}{\sqrt{2}}(|B\rangle+|D\rangle)$, the entanglement is spontaneously transferred from modes $B$ and $D$ to $A$ and $C$, which is attributed to that only the entangled state $|\nu_1^{(2)}\rangle$ will linearly grow over time and be the dominance of evolution as $t\to\infty$. Similarly, the entanglement between modes $A$ and $C$ can be also transfered to $B$ and $D$ spontaneously by choosing, such as, $|\Psi(0)\rangle=\frac{1}{\sqrt{2}}(|A\rangle+|C\rangle)$.

Combining these EP dynamics, we immediately become aware of an application described by the entanglement generation at a 3rd-order EP (e.g., $\epsilon\neq\pm i$, $\kappa=1$) and the subsequent transfer at two degenerate 2nd-order EPs (e.g., $\epsilon=-i$, $\kappa=-1$), which converts $|\nu_1\rangle$ to, for instance, $|\nu_1^{(2)}\rangle$, allowing the entanglement readout without altering its form. Figure~\ref{fig3}(c) shows this readout process demonstrated by the fidelities $F_{1}^{(1)}(t)$ and $F_{2}^{(1)}(t)$ for the normalized $|\Psi(t)\rangle$, where $F_{1}^{(1)}(t)=|\langle\nu_{1}'|\Psi(t)\rangle|^{2}$ with the initial state $|\nu_{1}'\rangle=|\nu_{1}\rangle/\sqrt{\langle\nu_{1}|\nu_{1}\rangle}$ and $F_{2}^{(1)}(t)=|\langle\nu_{1}^{(2)\prime}|\Psi(t)\rangle|^{2}$ with the target state $|\nu^{(2)\prime}_{1}\rangle=|\nu^{(2)}_{1}\rangle/\sqrt{\langle\nu^{(2)}_{1}|\nu^{(2)}_{1}\rangle}$. In this case, we can also read out $|\nu_{1}\rangle$ for $\epsilon=i$ and $\kappa\neq\pm1$, as indicated by the fidelities $F_{1}^{(2)}(t)$ and $F_{2}^{(2)}(t)$ in Fig.~\ref{fig3}(c), with $F_{1}^{(2)}(t)=|\langle\nu_{1}'|\Psi(t)\rangle|^{2}$ and $F_{2}^{(2)}(t)=|\langle\nu_{1}^{(1)\prime}|\Psi(t)\rangle|^{2}$.

\emph{Conclusion}.---In summary, we established PCRs to elucidate the EP dynamics. Despite PH-guaranteed real spectra, depending on EP order and initial condition, the long-term evolution is governed by the polynomially growing coalescing eigenstates or their superposition, responsible for a spontaneous state conversion. We also validated the dynamics by examining a NH stub ribbon with PH and a $\mathcal{PT}$-symmetric NH diamond ring, where the nonreciprocal hopping and the complex coupling can be implemented by auxiliary dissipative modes~\cite{SM}. These models exploit applications like spontaneous multiport beam splitter with arbitrary ratio and spontaneous entanglement generation and transfer. This work unravels the evolution of NH systems at EPs, paving the way for engineering new devices and technologies based on the PH-driven EP dynamics.

\emph{Acknowledgements}.---This work was supported by the National Natural Science Foundation of China (Grants No. 12347187, No. 12274376) and the China Postdoctoral Science Foundation (Grant No. 2023M743226).

\makeatletter
\let\oldaddcontentsline\addcontentsline
\renewcommand{\addcontentsline}[3]{} 

\let\addcontentsline\oldaddcontentsline 
\makeatother

\myemptypage
\pagebreak
\widetext
\begin{center}
	\textbf{\large Supplementary material for ``Exceptional-Point Dynamics''}
\end{center}

\setcounter{equation}{0}
\setcounter{figure}{0}
\setcounter{table}{0}
\setcounter{page}{1}
\makeatletter
\renewcommand{\theequation}{S\arabic{equation}}
\renewcommand{\thefigure}{S\arabic{figure}}
\renewcommand{\bibnumfmt}[1]{[S#1]}
\renewcommand{\citenumfont}[1]{S#1}

\newcommand\bblue[1]{\textcolor{blue}{\textbf{#1}}}
\newcommand\bgreen [1]{\textcolor{green}{\textbf{#1}}}
\newcommand\bred[1]{\textcolor{red}{\textbf{#1}}}
\newcommand\bblack[1]{\textcolor{black}{\textbf{#1}}}

\newcommand{\rmnum}[1]{\romannumeral #1}
\newcommand{\Rmnum}[1]{\expandafter\@slowromancap\romannumeral #1@}

\tableofcontents
\makeatother

\section{Pseudo-completeness relations at exceptional points}

\subsection{Pseudo-completeness relation for a single arbitrary-order exceptional point}
Let us embark on a general time-independent non-Hermitian (NH) Hamiltonian $\hat{H}(\lambda_{1},\dots,\lambda_{K})$ of dimension $M\times M$, with $\lambda_{1}$, $\dots$, $\lambda_{K}$ tunable system parameters. As the $K$ parameters vary, when $\hat{H}$ encounters an exceptional point (EP) of order $N$, viz., $N$ eigenenergies in the system spectrum become degenerate at $E_{\mathrm{EP}}$ and their eigenstates coalesce into one simultaneously, so that the spectral singularity and the skewness of eigenspace occur, the defective $\hat{H}$ is similar to a Jordan matrix $\hat{H}_{J}$, accompanied by the Jordan decomposition $\hat{S}_{\bot}^{-1}\hat{H}\hat{S}_{\bot}=\hat{H}_{J}$. $\hat{H}_{J}$ can be written in the form of $\hat{H}_{J}=\hat{J}\oplus\hat{D}$, where the Jordan block $\hat{J}$ has elements $\hat{J}_{n,n^{\prime}}=E_{\mathrm{EP}}\delta_{n,n^{\prime}}+\delta_{n,n^{\prime}-1}$ ($n,n^{\prime}\in[1,N]$), with $\delta$ the Kronecker symbol, and the elements of the diagonal $\hat{D}$ are $\hat{D}_{j,j^{\prime}}=E_{j}\delta_{j,j^{\prime}}$ ($j,j^{\prime}\in[1,M-N]$), with $E_{j}$ the remainder of the eigenenergies. Here, we assume nondegenerate $E_{j}$ and $E_{j}\neq E_{\mathrm{EP}}$. The invertible transformation matrix $\hat{S}_{\bot}=(|\psi_{1}\rangle,\dots,|\psi_{N}\rangle,|\phi_{1}\rangle,\dots,|\phi_{M-N}\rangle)$, with $|\psi_{1}\rangle$ the eigenstate of $E_{\mathrm{EP}}$, $|\psi_{n_{1}}\rangle$ ($n_{1}\in[2,N]$) the generalized eigenstates determined by the Jordan chain $(\hat{H}-E_{\mathrm{EP}})|\psi_{n_{1}}\rangle=|\psi_{n_{1}-1}\rangle$, and $|\phi_{j}\rangle$ the eigenstate of $E_{j}$. Recall that for a nondefective $\hat{H}$, its eigenstate $|R_{m}\rangle$ with eigenenergy $E_{m}$ and the corresponding eigenstate $|L_{m}\rangle$ with eigenenergy $E_{m}^{\ast}$ of $\hat{H}^{\dagger}$, commonly referred to as right and left eigenstates, together constitute a biorthogonal basis, satisfying the orthonormalization condition $\langle L_m|R_{m'}\rangle/(\sqrt{\langle L_m|R_m\rangle}\sqrt{\langle L_{m'}|R_{m'}\rangle})=\delta_{m,m'}$ ($m,m^{\prime}\in[1,M]$) and the completeness relation $\sum_{m=1}^M|R_m\rangle\langle L_m|/\langle L_m|R_m\rangle=\hat{\mathbb{I}}$, with $\hat{\mathbb{I}}$ the identity matrix. However, it is not the case once the defective degeneracy is present. Consider the Jordan decomposition $\hat{S}_{\top}^{-1}\hat{H}^{\dagger}\hat{S}_{\top}=\hat{H}_{J}^{\prime}$, where $\hat{H}_{J}^{\prime}=\hat{J}^{\prime}\oplus\hat{D}^{\prime}$, with $\hat{J}_{n,n^{\prime}}^{\prime}=E_{\mathrm{EP}}^{\ast}\delta_{n,n^{\prime}}+\delta_{n,n^{\prime}-1}$ and $\hat{D}_{j,j^{\prime}}^{\prime}=E_{j}^{\ast}\delta_{j,j^{\prime}}$, and $\hat{S}_{\top}=(|\tilde{\psi}_{1}\rangle,\dots,|\tilde{\psi}_{N}\rangle,|\tilde{\phi}_{1}\rangle,\dots,|\tilde{\phi}_{M-N}\rangle)$. While $\langle\tilde{\phi}_{j}|\psi_{1}\rangle=\langle\tilde{\psi}_{1}|\phi_{j}\rangle=0$ and the standard orthonormalization condition between $\langle\tilde{\phi}_{j}|$ and $|\phi_{j^{\prime}}\rangle$ persist, the self-orthogonality, $\langle\tilde{\psi}_{1}|\psi_{1}\rangle=0$, which roots in $\langle\tilde{\psi}_{2}|(\hat{H}-E_{\mathrm{EP}})|\psi_{1}\rangle=\langle\tilde{\psi}_{1}|(\hat{H}-E_{\mathrm{EP}})|\psi_{2}\rangle=0$, renders the inverse of the biorthogonal norm $\sqrt{\langle\tilde{\psi}_{1}|\psi_{1}\rangle}$ divergent, further making both the normalization of $\{\langle\tilde{\psi}_{1}|,|\psi_{1}\rangle\}$ and the completeness relation fail to hold. 

In addition, from $\langle\tilde{\phi}_{j}|\psi_{1}\rangle=0$ and $\langle\tilde{\phi}_{j}|(\hat{H}-E_{\mathrm{EP}})|\psi_{n_{1}}\rangle=(E_{j}-E_{\mathrm{EP}})\langle\tilde{\phi}_{j}|\psi_{n_{1}}\rangle=\langle\tilde{\phi}_{j}|\psi_{n_{1}-1}\rangle$ , we can obtain $\langle\tilde{\phi}_{j}|\psi_{n_{1}}\rangle=0$. Similarly, $\langle\tilde{\psi}_{n_{1}}|\phi_{j}\rangle=0$. For $|\psi_{1}\rangle$, we also have $\langle\tilde{\psi}_{n_{1}-1}|\psi_{1}\rangle=0$ due to $\langle\tilde{\psi}_{n_{1}}|(\hat{H}-E_{\mathrm{EP}})|\psi_{1}\rangle=0$, leaving $\langle\tilde{\psi}_{N}|\psi_{1}\rangle$ unknown, for $|\psi_{2}\rangle$, $\langle\tilde{\psi}_{n_{2}-1}|\psi_{2}\rangle=0$ ($n_{2}\in[2,N-1]$) due to $\langle\tilde{\psi}_{n_{2}}|(\hat{H}-E_{\mathrm{EP}})|\psi_{2}\rangle=\langle\tilde{\psi}_{n_{2}}|\psi_{1}\rangle=0$, leaving $\langle\tilde{\psi}_{N-1}|\psi_{2}\rangle$ and $\langle\tilde{\psi}_{N}|\psi_{2}\rangle$ unknown, and so on. In a word, $\langle\tilde{\psi}_{N+1-l}|\psi_{n_{1}-1}\rangle=0$ ($l\in[n_{1},N]$), leaving $\langle\tilde{\psi}_{N+1-p}|\psi_{n}\rangle$ ($p\in[1,n]$) unknown.

On the other hand, it is found that all $|\psi_{n}\rangle$ are the zero-energy eigenstates of the matrix $\hat{\mathbb{H}}=(\hat{H}-E_{\mathrm{EP}})^{N}$, $\hat{\mathbb{H}}|\psi_{n}\rangle=0$, and each $|\phi_{j}\rangle$ is also the eigenstate of $\hat{\mathbb{H}}$ with eigenenergy $(E_{j}-E_{\mathrm{EP}})^{N}$, $\hat{\mathbb{H}}|\phi_{j}\rangle=(E_{j}-E_{\mathrm{EP}})^{N}|\phi_{j}\rangle$. Additionally, $\hat{S}_{\bot}^{-1}\hat{\mathbb{H}}\hat{S}_{\bot}=(\hat{H}_{J}-E_{\mathrm{EP}})^{N}$ with $\hat{H}_{J}-E_{\mathrm{EP}}=\hat{J}_{\alpha}\oplus\hat{D}_{\alpha}$, where the elements of $\hat{J}_{\alpha}=\hat{J}-E_{\mathrm{EP}}$ are $(\hat{J}_{\alpha})_{n,n^{\prime}}=\delta_{n,n^{\prime}-1}$ and $\hat{D}_{\alpha}=\hat{D}-E_{\mathrm{EP}}$ has elements $(\hat{D}_{\alpha})_{j,j^{\prime}}=(E_{j}-E_{\mathrm{EP}})\delta_{j,j^{\prime}}$. It is worth mentioning that $\hat{J}_{\alpha}$ is a nilpotent, viz., $\hat{J}_{\alpha}^{N}=\hat{\mathbb{O}}$, and $(\hat{H}_{J}-E_{\mathrm{EP}})^{N}$ is thus a diagonal matrix $\hat{J}_{\beta}\oplus\hat{D}_{\beta}$, with elements $(\hat{J}_{\beta})_{n,n^{\prime}}=0$ and $(\hat{D}_{\beta})_{j,j^{\prime}}=(E_{j}-E_{\mathrm{EP}})^{N}\delta_{j,j^{\prime}}$, which implies that the NH $\hat{\mathbb{H}}$ is a diagonalizable matrix with $N$ degenerate zero eigenenergies. Similarly, via $\hat{S}_{\top}$, the NH $\hat{\mathbb{H}}^{\dagger}$ can be diagonalized as $(\hat{H}_{J}^{\prime}-E_{\mathrm{EP}}^{\ast})^{N}=\hat{J}_{\beta}^{\prime}\oplus\hat{D}_{\beta}^{\prime}$, with elements $(\hat{J}_{\beta}^{\prime})_{n,n^{\prime}}=0$ and $(\hat{D}_{\beta}^{\prime})_{j,j^{\prime}}=(E_{j}^{\ast}-E_{\mathrm{EP}}^{\ast})^{N}\delta_{j,j^{\prime}}$, whose zero eigenenergy is also $N$-fold degenerate. Accordingly, in terms of $\hat{\mathbb{H}}$ and $\hat{\mathbb{H}}^{\dagger}$ with the nondefective degeneracy, their eigenstates can together constitute a biorthogonal basis. Based on the previous analysis and discussion, e.g., for $|\psi_{1}\rangle$, its left partner must be $\langle\tilde{\psi}_{N}|$, which stems from $\langle\tilde{\phi}_{j}|\psi_{1}\rangle=\langle\tilde{\psi}_{n_{1}-1}|\psi_{1}\rangle=0$ and leads to $\langle\tilde{\psi}_{N}|\psi_{1}\rangle\neq0$, for $|\psi_{2}\rangle$, its left partner must be $\langle\tilde{\psi}_{N-1}|$ since $\langle\tilde{\phi}_{j}|\psi_{2}\rangle=\langle\tilde{\psi}_{n_{2}-1}|\psi_{2}\rangle=0$ and $\langle\tilde{\psi}_{N}|$ is already matched with $|\psi_{1}\rangle$, giving rise to $\langle\tilde{\psi}_{N-1}|\psi_{2}\rangle\neq0$ and requiring $\langle\tilde{\psi}_{N}|\psi_{2}\rangle=0$, and so on. Therefore, we can draw a conclusion that the left partner of $|\psi_{n}\rangle$ must be $\langle\tilde{\psi}_{N+1-n}|$, viz., $\langle\tilde{\psi}_{N+1-n}|\psi_{n}\rangle\neq0$, and $\langle\tilde{\psi}_{N+1-q}|\psi_{n_{1}}\rangle$ ($q\in[1,n_{1}-1]$) need to be vanishing in the meantime. Note that if $\hat{S}_{\top}$ is rewritten in the form of $\hat{S}_{\top}^{\prime}=(|\tilde{\psi}_{N}\rangle,\dots,|\tilde{\psi}_{1}\rangle,|\tilde{\phi}_{1}\rangle,\dots,|\tilde{\phi}_{M-N}\rangle)$, the diagonal elements of $(\hat{H}_{J}^{\prime}-E_{\mathrm{EP}}^{\ast})^{N}$ will not change. Ultimately, we can redefine a set of left and right eigenstates subjected to the orthonormalization condition as 
\begin{equation}\label{SEq-1}
\begin{split}
\langle\tilde{\nu}_{N}|&=\frac{\langle\tilde{\psi}_{N}|}{\sqrt{\langle\tilde{\psi}_{N}|\psi_{1}\rangle}},\\
|\nu_{1}\rangle&=\frac{|\psi_{1}\rangle}{\sqrt{\langle\tilde{\psi}_{N}|\psi_{1}\rangle}},\\
\langle\tilde{\nu}_{N+1-n_{1}}|&=\frac{\langle\tilde{\mu}_{N+1-n_{1}}|}{\sqrt{\langle\tilde{\mu}_{N+1-n_{1}}|\mu_{n_{1}}\rangle}}=\frac{\langle\tilde{\psi}_{N+1-n_{1}}|-\sum_{q=1}^{n_{1}-1}\langle\tilde{\psi}_{N+1-n_{1}}|\nu_{q}\rangle\langle\tilde{\nu}_{N+1-q}|}{\sqrt{\langle\tilde{\mu}_{N+1-n_{1}}|\mu_{n_{1}}\rangle}}=\frac{\langle\tilde{\psi}_{N+1-n_{1}}|}{\sqrt{\langle\tilde{\psi}_{N+1-n_{1}}|\mu_{n_{1}}\rangle}},\\
|\nu_{n_{1}}\rangle&=\frac{|\mu_{n_{1}}\rangle}{\sqrt{\langle\tilde{\mu}_{N+1-n_{1}}|\mu_{n_{1}}\rangle}}=\frac{|\psi_{n_{1}}\rangle-\sum_{q=1}^{n_{1}-1}\langle\tilde{\nu}_{N+1-q}|\psi_{n_{1}}\rangle|\nu_{q}\rangle}{\sqrt{\langle\tilde{\psi}_{N+1-n_{1}}|\mu_{n_{1}}\rangle}},\\
\langle\tilde{\omega}_{j}|&=\frac{\langle\tilde{\phi}_{j}|}{\sqrt{\langle\tilde{\phi}_{j}|\phi_{j}\rangle}},\\
|\omega_{j}\rangle&=\frac{|\phi_{j}\rangle}{\sqrt{\langle\tilde{\phi}_{j}|\phi_{j}\rangle}},
\end{split}
\end{equation}
where $\langle\tilde{\psi}_{N+1-n_{1}}|\nu_{q}\rangle=0$ is employed. Moreover, $\langle\tilde{\psi}_{N+1-n_{1}}|\mu_{n_{1}}\rangle=\langle\tilde{\psi}_{N+1-n_{1}}|\psi_{n_{1}}\rangle=\langle\tilde{\psi}_{N}|(\hat{H}-E_{\mathrm{EP}})^{n_{1}-1}|\psi_{n_{1}}\rangle=\langle\tilde{\psi}_{N}|\psi_{1}\rangle=C$, and Eq.~(\ref{SEq-1}) can be thereby simplified as    
\begin{equation}\label{SEq-2}
\begin{aligned}
\langle\tilde{\nu}_{N+1-n}|&=\frac{\langle\tilde{\psi}_{N+1-n}|}{\sqrt{C}},&|\nu_{n}\rangle&=\frac{|\psi_{n}\rangle-\sum_{q=1}^{n-1}\langle\tilde{\nu}_{N+1-q}|\psi_{n}\rangle|\nu_{q}\rangle}{\sqrt{C}},\\
\langle\tilde{\omega}_{j}|&=\frac{\langle\tilde{\phi}_{j}|}{\sqrt{\langle\tilde{\phi}_{j}|\phi_{j}\rangle}},&|\omega_{j}\rangle&=\frac{|\phi_{j}\rangle}{\sqrt{\langle\tilde{\phi}_{j}|\phi_{j}\rangle}},
\end{aligned}
\end{equation}	
with $\langle\tilde{\nu}_{n}|\nu_{n^{\prime}}\rangle=\delta_{n,N+1-n^{\prime}}$, $\langle\tilde{\omega}_{j}|\omega_{j^{\prime}}\rangle=\delta_{j,j^{\prime}}$, and $\langle\tilde{\omega}_{j}|\nu_{n}\rangle=\langle\tilde{\nu}_{n}|\omega_{j}\rangle=0$, further enabling the following closure relation 
\begin{equation}\label{SEq-3}
\begin{split}
\sum_{n=1}^{N}|\nu_{n}\rangle\langle\tilde{\nu}_{N+1-n}|+\sum_{j=1}^{M-N}|\omega_{j}\rangle\langle\tilde{\omega}_{j}|=\hat{\mathbb{I}},
\end{split}
\end{equation}
which we deem as the pseudo-completeness relation (PCR) in the context of $\hat{H}$. Actually, equation~(\ref{SEq-3}) can be demonstrated as follows: In the new biorthogonal basis, $\hat{S}_{\bot}=(|\nu_{1}\rangle,\dots,|\nu_{N}\rangle,|\omega_{1}\rangle,\dots,|\omega_{M-N}\rangle)$, $\hat{S}_{\top}^{\prime}=(|\tilde{\nu}_{N}\rangle,\dots,|\tilde{\nu}_{1}\rangle,|\tilde{\omega}_{1}\rangle,\dots,|\tilde{\omega}_{M-N}\rangle)$, and we can reformulate
\begin{equation}\label{SEq-4}
\begin{aligned}
\langle\tilde{\nu}_{N}|&=(1,0,\dots,0,0,0,\dots,0)_{1\times M}(\hat{S}_{\top}^{\prime})^{\dagger},&|\nu_{1}\rangle&=\hat{S}_{\bot}(1,0,\dots,0,0,0,\dots,0)_{1\times M}^{\mathrm{T}},\\
\langle\tilde{\nu}_{N-1}|&=(0,1,\dots,0,0,0,\dots,0)_{1\times M}(\hat{S}_{\top}^{\prime})^{\dagger},&|\nu_{2}\rangle&=\hat{S}_{\bot}(0,1,\dots,0,0,0,\dots,0)_{1\times M}^{\mathrm{T}},\\
\vdots&&\vdots&\\
\langle\tilde{\nu}_{1}|&=(0,0,\dots,1,0,0,\dots,0)_{1\times M}(\hat{S}_{\top}^{\prime})^{\dagger},&|\nu_{N}\rangle&=\hat{S}_{\bot}(0,0,\dots,1,0,0,\dots,0)_{1\times M}^{\mathrm{T}},\\
\langle\tilde{\omega}_{1}|&=(0,0,\dots,0,1,0,\dots,0)_{1\times M}(\hat{S}_{\top}^{\prime})^{\dagger},&|\omega_{1}\rangle&=\hat{S}_{\bot}(0,0,\dots,0,1,0,\dots,0)_{1\times M}^{\mathrm{T}},\\
\langle\tilde{\omega}_{2}|&=(0,0,\dots,0,0,1,\dots,0)_{1\times M}(\hat{S}_{\top}^{\prime})^{\dagger},&|\omega_{2}\rangle&=\hat{S}_{\bot}(0,0,\dots,0,0,1,\dots,0)_{1\times M}^{\mathrm{T}},\\
\vdots&&\vdots&\\
\langle\tilde{\omega}_{M-N}|&=(0,0,\dots,0,0,0,\dots,1)_{1\times M}(\hat{S}_{\top}^{\prime})^{\dagger},&|\omega_{M-N}\rangle&=\hat{S}_{\bot}(0,0,\dots,0,0,0,\dots,1)_{1\times M}^{\mathrm{T}},
\end{aligned}
\end{equation}    
with $\mathrm{T}$ the transpose operation, so that the left-hand side of Eq.~(\ref{SEq-3}) is converted into $\hat{S}_{\bot}(\hat{S}_{\top}^{\prime})^{\dagger}$. Furthermore, what we have learned is  $(\hat{S}_{\top}^{\prime})^{\dagger}\hat{S}_{\bot}=\hat{\mathbb{I}}$ by resorting to the orthonormalization condition, yielding $\hat{S}_{\bot}^{-1}[(\hat{S}_{\top}^{\prime})^{\dagger}]^{-1}=\hat{\mathbb{I}}\rightarrow\hat{\mathbb{I}}=\hat{S}_{\bot}(\hat{S}_{\top}^{\prime})^{\dagger}$. Eventually, we arrive at Eq.~(\ref{SEq-3}).

Unambiguously, all $\langle\tilde{\nu}_{n}|$ still obey the Jordan chain, as well as $|\nu_{n}\rangle$,
\begin{equation}\label{SEq-5}
\begin{split}
(\hat{H}-E_{\mathrm{EP}})|\nu_{n_{1}}\rangle=\frac{|\psi_{n_{1}-1}\rangle-\sum_{q=1}^{n_{1}-2}\langle\tilde{\nu}_{N-q}|\psi_{n_{1}}\rangle|\nu_{q}\rangle}{\sqrt{C}}=\frac{|\psi_{n_{1}-1}\rangle-\sum_{q=1}^{n_{1}-2}\langle\tilde{\nu}_{N+1-q}|\psi_{n_{1}-1}\rangle|\nu_{q}\rangle}{\sqrt{C}}=|\nu_{n_{1}-1}\rangle.
\end{split}
\end{equation}

\subsection{Pseudo-completeness relation for multiple arbitrary-order exceptional points without degeneracy}
Having investigated the PCR of the time-independent NH $\hat{H}$ with only one $N$th-order EP, we now turn to a more general case where multiple EPs are present. Consider a $M\times M$ $\hat{\mathcal{H}}$ encountering simultaneously $\mathbb{N}$ EPs, with respective coalescence eigenenergies $E_{\mathrm{EP}}^{(s)}$ ($s\in[1,\mathbb{N}]$), and the order of each EP is given by $N_{s}$. The Jordan canonical form of $\hat{\mathcal{H}}$ reads $\hat{\mathcal{H}}_{J}=(\oplus_{s=1}^{\mathbb{N}}\hat{\mathcal{J}}_{s})\oplus\hat{\mathcal{D}}$, with $\hat{\mathcal{J}}_{s}$ the Jordan block of dimension $N_{s}\times N_{s}$ of $E_{\mathrm{EP}}^{(s)}$ and $\hat{\mathcal{D}}$ the diagonal matrix consisting of the rest of the eigenenergies $E_{r}$ ($r\in[1,M-\sum_{s=1}^{\mathbb{N}}N_{s}]$). Firstly, we still assume that all $E_{\mathrm{EP}}^{(s)}$ and $E_{r}$ are nondegenerate. The similarity transformation matrix $\hat{\mathcal{S}}_{\bot}=(\dots,|\psi_{1}^{(s)}\rangle,\dots,|\psi_{N_{s}}^{(s)}\rangle,\dots,|\phi_{r}\rangle,\dots)$, with $|\psi_{1}^{(s)}\rangle$ the eigenstate of $E_{\mathrm{EP}}^{(s)}$, $|\psi_{g_{s}}^{(s)}\rangle$ ($g_{s}\in[2,N_{s}]$) the corresponding generalized eigenstates, and $|\phi_{r}\rangle$ the eigenstate of $E_{r}$. Also, for $\hat{\mathcal{H}}^{\dagger}$, we have the eigenstate and generalized eigenstates of $E_{\mathrm{EP}}^{(s)\ast}$, $|\tilde{\psi}_{1}^{(s)}\rangle$ and $|\tilde{\psi}_{g_{s}}^{(s)}\rangle$, and the eigenstate of $E_{r}^{\ast}$, $|\tilde{\phi}_{r}\rangle$.

In order to establish the so-called PCR, the biorthogonality among these eigenstates and generalized eigenstates needs to be analyzed and discussed. We can now examine that $\langle\tilde{\phi}_{r}|\phi_{r^{\prime}}\rangle=0$ ($r\neq r^{\prime}$), $\langle\tilde{\phi}_{r}|\psi_{1}^{(s)}\rangle=\langle\tilde{\phi}_{r}|\psi_{g_{s}}^{(s)}\rangle=\langle\tilde{\psi}_{1}^{(s)}|\phi_{r}\rangle=\langle\tilde{\psi}_{g_{s}}^{(s)}|\phi_{r}\rangle=0$, and $\langle\tilde{\psi}_{1}^{(s)}|\psi_{1}^{(s^{\prime})}\rangle=\langle\tilde{\psi}_{1}^{(s)}|\psi_{g_{s^{\prime}}}^{(s^{\prime})}\rangle=\langle\tilde{\psi}_{g_{s}}^{(s)}|\psi_{1}^{(s^{\prime})}\rangle=0$ ($s\neq s^{\prime}$), further leading to $\langle\tilde{\psi}_{g_{s}}^{(s)}|\psi_{g_{s^{\prime}}}^{(s^{\prime})}\rangle=0$, which can be clarified as follows: We first have $\langle\tilde{\psi}_{g_{s}}^{(s)}|(\hat{\mathcal{H}}-E_{\mathrm{EP}}^{(s^{\prime})})|\psi_{g_{s^{\prime}}}^{(s^{\prime})}\rangle=\langle\tilde{\psi}_{g_{s}}^{(s)}|\psi_{g_{s^{\prime}}-1}^{(s^{\prime})}\rangle$, then, $\langle\tilde{\psi}_{g_{s}}^{(s)}|(\hat{\mathcal{H}}-E_{\mathrm{EP}}^{(s)}+E_{\mathrm{EP}}^{(s)}-E_{\mathrm{EP}}^{(s^{\prime})})|\psi_{g_{s^{\prime}}}^{(s^{\prime})}\rangle=\langle\tilde{\psi}_{g_{s}}^{(s)}|\psi_{g_{s^{\prime}}-1}^{(s^{\prime})}\rangle$, finally, we obtain the recurrence relation $\langle\tilde{\psi}_{g_{s}-1}^{(s)}|\psi_{g_{s^{\prime}}}^{(s^{\prime})}\rangle+(E_{\mathrm{EP}}^{(s)}-E_{\mathrm{EP}}^{(s^{\prime})})\langle\tilde{\psi}_{g_{s}}^{(s)}|\psi_{g_{s^{\prime}}}^{(s^{\prime})}\rangle=\langle\tilde{\psi}_{g_{s}}^{(s)}|\psi_{g_{s^{\prime}}-1}^{(s^{\prime})}\rangle$. With the initial conditions $\langle\tilde{\psi}_{1}^{(s)}|\psi_{g_{s^{\prime}}}^{(s^{\prime})}\rangle=\langle\tilde{\psi}_{g_{s}}^{(s)}|\psi_{1}^{(s^{\prime})}\rangle=0$, we can find $\langle\tilde{\psi}_{g_{s}-1}^{(s)}|\psi_{g_{s^{\prime}}}^{(s^{\prime})}\rangle=\langle\tilde{\psi}_{g_{s}}^{(s)}|\psi_{g_{s^{\prime}}-1}^{(s^{\prime})}\rangle=0$, so that $\langle\tilde{\psi}_{g_{s}}^{(s)}|\psi_{g_{s^{\prime}}}^{(s^{\prime})}\rangle$ must be vanishing. In addition, for each $E_{\mathrm{EP}}^{(s)}$, its eigenstate and generalized eigenstates still satisfy $\langle\tilde{\psi}_{N_{s}+1-x_{s}}^{(s)}|\psi_{g_{s}-1}^{(s)}\rangle=0$ ($x_{s}\in[g_{s},N_{s}]$), with unknown $\langle\tilde{\psi}_{N_{s}+1-z_{s}}^{(s)}|\psi_{y_{s}}^{(s)}\rangle$ ($y_{s}\in[1,N_{s}],~z_{s}\in[1,y_{s}]$), coinciding with the case of a single EP. Similarly, it turns out that all $|\psi_{y_{s}}^{(s)}\rangle$ are the eigenstates of the matrix $\hat{\mathcal{A}}_{\mathbb{N}}$ with a $N_{s}$-fold degenerate eigenenergy $\mathbb{E}_{\mathrm{DP}}^{(s)}=(\cdots((-a_{s})^{N_{s+1}}-a_{s+1})^{N_{s+2}}\cdots-a_{\mathbb{N}-1})^{N_{\mathbb{N}}}$ and $\mathbb{E}_{\mathrm{DP}}^{(\mathbb{N})}=0$, where $\hat{\mathcal{A}}_{s}=(\hat{\mathcal{A}}_{s-1}-a_{s-1})^{N_{s}}$, with $\hat{\mathcal{A}}_{0}=\hat{\mathcal{H}}$, and $a_{s}$ is defined as $a_{s}=(\cdots((E_{\mathrm{EP}}^{(s+1)}-a_{0})^{N_{1}}-a_{1})^{N_{2}}\cdots-a_{s-1})^{N_{s}}$, with $a_{0}=E_{\mathrm{EP}}^{(1)}$, and each $|\phi_{r}\rangle$ is also the eigenstate of $\hat{\mathcal{A}}_{\mathbb{N}}$ with eigenenergy $\mathbb{E}_{r}=(\cdots((E_{r}-a_{0})^{N_{1}}-a_{1})^{N_{2}}\cdots-a_{\mathbb{N}-1})^{N_{\mathbb{N}}}$. It can be validated by considering a matrix $\hat{\mathbb{A}}_{\mathbb{N}}$, with  $\hat{\mathbb{A}}_{s}=(\hat{\mathbb{A}}_{s-1}-a_{s-1})^{N_{s}}$ and $\hat{\mathbb{A}}_{0}=\hat{\mathcal{H}}_{J}$, we can find that $\hat{\mathbb{A}}_{\mathbb{N}}$ is diagonal, $\hat{\mathbb{A}}_{\mathbb{N}}=\operatorname{diag}(\underbrace{\mathbb{E}_{\mathrm{DP}}^{(1)},\dots,\mathbb{E}_{\mathrm{DP}}^{(1)}}_{N_{1}},\underbrace{\mathbb{E}_{\mathrm{DP}}^{(2)},\dots,\mathbb{E}_{\mathrm{DP}}^{(2)}}_{N_{2}},\dots,\underbrace{\mathbb{E}_{\mathrm{DP}}^{(\mathbb{N})},\dots,\mathbb{E}_{\mathrm{DP}}^{(\mathbb{N})}}_{N_{\mathbb{N}}},\mathbb{E}_{1},\dots,\mathbb{E}_{M-\sum_{s=1}^{\mathbb{N}}N_{s}})$, and $\hat{\mathcal{S}}_{\bot}\hat{\mathbb{A}}_{\mathbb{N}}\hat{\mathcal{S}}_{\bot}^{-1}=\hat{\mathcal{A}}_{\mathbb{N}}$, which suggests that each eigenstate of $\hat{\mathcal{A}}_{\mathbb{N}}$ can be expressed, respectively, as
\begin{equation}\label{SEq-6}
\begin{split}
\hat{\mathcal{S}}_{\bot}(1,0,\dots,0,0,0,\dots,0,\dots,0,0,\dots,0,0,0,\dots,0)_{1\times M}^{\mathrm{T}}&=|\psi_{1}^{(1)}\rangle,\\
\hat{\mathcal{S}}_{\bot}(0,1,\dots,0,0,0,\dots,0,\dots,0,0,\dots,0,0,0,\dots,0)_{1\times M}^{\mathrm{T}}&=|\psi_{2}^{(1)}\rangle,\\
&\vdots\\
\hat{\mathcal{S}}_{\bot}(0,0,\dots,1,0,0,\dots,0,\dots,0,0,\dots,0,0,0,\dots,0)_{1\times M}^{\mathrm{T}}&=|\psi_{N_{1}}^{(1)}\rangle,\\
\hat{\mathcal{S}}_{\bot}(0,0,\dots,0,1,0,\dots,0,\dots,0,0,\dots,0,0,0,\dots,0)_{1\times M}^{\mathrm{T}}&=|\psi_{1}^{(2)}\rangle,\\
\hat{\mathcal{S}}_{\bot}(0,0,\dots,0,0,1,\dots,0,\dots,0,0,\dots,0,0,0,\dots,0)_{1\times M}^{\mathrm{T}}&=|\psi_{2}^{(2)}\rangle,\\
&\vdots\\
\hat{\mathcal{S}}_{\bot}(0,0,\dots,0,0,0,\dots,1,\dots,0,0,\dots,0,0,0,\dots,0)_{1\times M}^{\mathrm{T}}&=|\psi_{N_{2}}^{(2)}\rangle,\\
&\vdots\\
\hat{\mathcal{S}}_{\bot}(0,0,\dots,0,0,0,\dots,0,\dots,1,0,\dots,0,0,0,\dots,0)_{1\times M}^{\mathrm{T}}&=|\psi_{1}^{(\mathbb{N})}\rangle,\\
\hat{\mathcal{S}}_{\bot}(0,0,\dots,0,0,0,\dots,0,\dots,0,1,\dots,0,0,0,\dots,0)_{1\times M}^{\mathrm{T}}&=|\psi_{2}^{(\mathbb{N})}\rangle,\\
&\vdots\\
\hat{\mathcal{S}}_{\bot}(0,0,\dots,0,0,0,\dots,0,\dots,0,0,\dots,1,0,0,\dots,0)_{1\times M}^{\mathrm{T}}&=|\psi_{N_{\mathbb{N}}}^{(\mathbb{N})}\rangle,\\
\hat{\mathcal{S}}_{\bot}(0,0,\dots,0,0,0,\dots,0,\dots,0,0,\dots,0,1,0,\dots,0)_{1\times M}^{\mathrm{T}}&=|\phi_{1}\rangle,\\
\hat{\mathcal{S}}_{\bot}(0,0,\dots,0,0,0,\dots,0,\dots,0,0,\dots,0,0,1,\dots,0)_{1\times M}^{\mathrm{T}}&=|\phi_{2}\rangle,\\
&\vdots\\
\hat{\mathcal{S}}_{\bot}(0,0,\dots,0,0,0,\dots,0,\dots,0,0,\dots,0,0,0,\dots,1)_{1\times M}^{\mathrm{T}}&=|\phi_{M-\sum_{s=1}^{\mathbb{N}}N_{s}}\rangle,
\end{split}
\end{equation}   
and has the corresponding eigenenergy mentioned above. For $|\tilde{\psi}_{y_{s}}^{(s)}\rangle$ and $|\tilde{\phi}_{r}\rangle$, we can also demonstrate with the same method that they are the eigenstates of $\hat{\mathcal{A}}_{\mathbb{N}}^{\dagger}$ with eigenenergies $\mathbb{E}_{\mathrm{DP}}^{(s)\ast}$ and $\mathbb{E}_{r}^{\ast}$. Consequently, in the context of $\hat{\mathcal{A}}_{\mathbb{N}}$ and $\hat{\mathcal{A}}_{\mathbb{N}}^{\dagger}$, we can further constitute a biorthogonal basis by resorting to $|\psi_{y_{s}}^{(s)}\rangle$, $|\tilde{\psi}_{y_{s}}^{(s)}\rangle$, $|\phi_{r}\rangle$ and $|\tilde{\phi}_{r}\rangle$. Combined with the previous analysis of biorthogonality, for each $E_{\mathrm{EP}}^{(s)}$, the left partner of $|\psi_{y_{s}}^{(s)}\rangle$ must be $\langle\tilde{\psi}_{N_{s}+1-y_{s}}^{(s)}|$, requiring $\langle\tilde{\psi}_{N_{s}+1-y_{s}}^{(s)}|\psi_{y_{s}}^{(s)}\rangle\neq0$ and $\langle\tilde{\psi}_{N_{s}+1-f}^{(s)}|\psi_{g_{s}}\rangle=0$ ($f\in[1,g_{s}-1])$ at the same time, which will next return to the discussion of a single EP case. Eventually, similar to Eq.~(\ref{SEq-2}), we can redefine a set of left and right eigenstates subjected to the orthonormalization condition as 
\begin{equation}\label{SEq-7}
\begin{aligned}
\langle\tilde{\nu}_{N_{s}+1-y_{s}}^{(s)}|&=\frac{\langle\tilde{\psi}_{N_{s}+1-y_{s}}^{(s)}|}{\sqrt{C_{s}}},&|\nu_{y_{s}}^{(s)}\rangle&=\frac{|\psi_{y_{s}}^{(s)}\rangle-\sum_{f=1}^{y_{s}-1}\langle\tilde{\nu}_{N_{s}+1-f}^{(s)}|\psi_{y_{s}}^{(s)}\rangle|\nu_{f}^{(s)}\rangle}{\sqrt{C_{s}}},\\
\langle\tilde{\omega}_{r}|&=\frac{\langle\tilde{\phi}_{r}|}{\sqrt{\langle\tilde{\phi}_{r}|\phi_{r}\rangle}},&|\omega_{r}\rangle&=\frac{|\phi_{r}\rangle}{\sqrt{\langle\tilde{\phi}_{r}|\phi_{r}\rangle}},
\end{aligned}
\end{equation}
where $C_{s}=\langle\tilde{\psi}_{N_{s}}^{(s)}|\psi_{1}^{(s)}\rangle$. Clearly, $\langle \tilde{\nu}_{y_s}^{(s)}|\nu_{y_{s^{\prime}}^{\prime}}^{(s^{\prime})}\rangle=\delta_{s,s^{\prime}}\delta_{y_s,N_s+1-y_{s^{\prime}}^{\prime}}$, $\langle\tilde{\omega}_{r}|\omega_{r^{\prime}}\rangle=\delta_{r,r^{\prime}}$, and $\langle\tilde{\omega}_{r}|\nu_{y_{s}}^{(s)}\rangle=\langle\tilde{\nu}_{y_{s}}^{(s)}|\omega_{r}\rangle=0$. The PCR of $\hat{\mathcal{H}}$ can be thus written as 
\begin{equation}\label{SEq-8}
\begin{split}
\sum_{s=1}^{\mathbb{N}}\sum_{y_{s}=1}^{N_{s}}|\nu_{y_{s}}^{(s)}\rangle\langle\tilde{\nu}_{N_{s}+1-y_{s}}^{(s)}|+\sum_{r=1}^{\mathbb{M}}|\omega_{r}\rangle\langle\tilde{\omega}_{r}|=\hat{\mathbb{I}},
\end{split}
\end{equation} 
with $\mathbb{M}=M-\sum_{s=1}^{\mathbb{N}}N_{s}$. Note that Eq.~(\ref{SEq-8}) reduces to Eq.~(\ref{SEq-3}) once $\mathbb{N}=1$.

\subsection{Pseudo-completeness relation for multiple arbitrary-order exceptional points with degeneracy}
For these $\mathbb{N}$ EPs, when there are $\mathbb{L}$ degenerate cases, and each has $L_{d}$ ($d\in[1,\mathbb{L}]$) degenerate eigenenergies, for the convenience of later description and without loss of generality, we successively arrange in each case the $L_{d}$ degenerate eigenenergies in descending order according to their algebraic multiplicities, so that $E_{\mathrm{EP}}^{(s)}=E_{\mathrm{EP}}^{(s+1)}=\dots=E_{\mathrm{EP}}^{(s+L_{d}-1)}$ with $N_{s}>N_{s+1}>\dots>N_{s+L_{d}-1}$. We can now find that all $|\psi_{y_{s+k}}^{(s+k)}\rangle$ ($k\in[0,L_{d}-1]$) are the eigenstates of $\hat{\mathcal{A}}_{\mathbb{N}}$ with a $(\sum_{k=0}^{L_{d}-1}N_{s+k})$-fold degenerate eigenenergy $\mathbb{E}_{\mathrm{DP}}^{(s+L_{d}-1)}=(\cdots((-a_{s+L_{d}-1})^{N_{s+L_{d}}}-a_{s+L_{d}})^{N_{s+L_{d}+1}}\cdots-a_{\mathbb{N}-1})^{N_{\mathbb{N}}}$ due to $a_{s}=a_{s+1}=\dots=a_{s+L_{d}-2}=0$, and all $|\tilde{\psi}_{y_{s+k}}^{(s+k)}\rangle$ are the eigenstates of $\hat{\mathcal{A}}_{\mathbb{N}}^{\dagger}$ with a $(\sum_{k=0}^{L_{d}-1}N_{s+k})$-fold degenerate eigenenergy $\mathbb{E}_{\mathrm{DP}}^{(s+L_{d}-1)\ast}$. Accordingly, while the biorthogonality among the $\mathbb{L}$ degenerate cases and that between different degenerate cases and the remaining $(\mathbb{N}-\sum_{d=1}^{\mathbb{L}}L_{d})$ nondegenerate cases persist, for each degenerate case, the biorthogonality between $\langle\tilde{\psi}_{y_{s+k}}^{(s+k)}|$ and $|\psi_{y_{s+k^{\prime}}}^{(s+k^{\prime})}\rangle$ should be guaranteed, which can be achieved via the following process:

(\romannumeral1) We redefine $\langle\tilde{\psi}_{y_{s}}^{(s)}|$ and $|\psi_{y_{s}}^{(s)}\rangle$ as $\langle\tilde{\nu}_{y_{s}}^{(s)}|$ and $|\nu_{y_{s}}^{(s)}\rangle$, respectively, by Eq.~(\ref{SEq-7});

(\romannumeral2) Since $\langle\tilde{\psi}_{y_{s+1}}^{(s+1)}|\nu_{b_{s|s+1}}^{(s)}\rangle=\langle\tilde{\nu}_{b_{s|s+1}}^{(s)}|\psi_{y_{s+1}}^{(s+1)}\rangle=0$ ($b_{s|s+1}\in[1,N_{s}-y_{s+1}]$), we first rewrite $\langle\tilde{\psi}_{y_{s+1}}^{(s+1)}|$ and $|\psi_{y_{s+1}}^{(s+1)}\rangle$ as
\begin{equation}\label{SEq-9}
\begin{split}
\langle\tilde{\upsilon}_{y_{s+1}}^{(s+1)}|&=\langle\tilde{\psi}_{y_{s+1}}^{(s+1)}|-\sum_{z_{s+1}=1}^{y_{s+1}}\langle\tilde{\psi}_{y_{s+1}}^{(s+1)}|\nu_{N_{s}+1-z_{s+1}}^{(s)}\rangle\langle\tilde{\nu}_{z_{s+1}}^{(s)}|,\\
|\upsilon_{y_{s+1}}^{(s+1)}\rangle&=|\psi_{y_{s+1}}^{(s+1)}\rangle-\sum_{z_{s+1}=1}^{y_{s+1}}\langle\tilde{\nu}_{N_{s}+1-z_{s+1}}^{(s)}|\psi_{y_{s+1}}^{(s+1)}\rangle|\nu_{z_{s+1}}^{(s)}\rangle,
\end{split}
\end{equation}
where $\langle\tilde{\upsilon}_{y_{s+1}}^{(s+1)}|$ and $|\upsilon_{y_{s+1}}^{(s+1)}\rangle$ also obey the Jordan chain and $\langle\tilde{\upsilon}_{y_{s+1}}^{(s+1)}|\nu_{y_{s}}^{(s)}\rangle=\langle\tilde{\nu}_{y_{s}}^{(s)}|\upsilon_{y_{s+1}}^{(s+1)}\rangle=0$. Subsequently, we redefine $\langle\tilde{\upsilon}_{y_{s+1}}^{(s+1)}|$ and $|\upsilon_{y_{s+1}}^{(s+1)}\rangle$ as $\langle\tilde{\nu}_{y_{s+1}}^{(s+1)}|$ and $|\nu_{y_{s+1}}^{(s+1)}\rangle$ by Eq.~(\ref{SEq-7}), with $C_{s+1}=\langle\tilde{\upsilon}_{N_{s+1}}^{(s+1)}|\upsilon_{1}^{(s+1)}\rangle=\langle\tilde{\psi}_{N_{s+1}}^{(s+1)}|\psi_{1}^{(s+1)}\rangle$;

(\romannumeral3) In the same way, since $\langle\tilde{\psi}_{y_{s+2}}^{(s+2)}|\nu_{b_{s|s+2}}^{(s)}\rangle=\langle\tilde{\nu}_{b_{s|s+2}}^{(s)}|\psi_{y_{s+2}}^{(s+2)}\rangle=0$ ($b_{s|s+2}\in[1,N_{s}-y_{s+2}]$) and $\langle\tilde{\psi}_{y_{s+2}}^{(s+2)}|\nu_{b_{s+1|s+2}}^{(s+1)}\rangle=\langle\tilde{\nu}_{b_{s+1|s+2}}^{(s+1)}|\psi_{y_{s+2}}^{(s+2)}\rangle=0$ ($b_{s+1|s+2}\in[1,N_{s+1}-y_{s+2}]$), we rewrite $\langle\tilde{\psi}_{y_{s+2}}^{(s+2)}|$ and $|\psi_{y_{s+2}}^{(s+2)}\rangle$ as
\begin{equation}\label{SEq-10}
\begin{split}
\langle\tilde{\upsilon}_{y_{s+2}}^{(s+2)}|&=\langle\tilde{\psi}_{y_{s+2}}^{(s+2)}|-\sum_{z_{s+2}=1}^{y_{s+2}}\langle\tilde{\psi}_{y_{s+2}}^{(s+2)}|\nu_{N_{s}+1-z_{s+2}}^{(s)}\rangle\langle\tilde{\nu}_{z_{s+2}}^{(s)}|-\sum_{z_{s+2}=1}^{y_{s+2}}\langle\tilde{\psi}_{y_{s+2}}^{(s+2)}|\nu_{N_{s+1}+1-z_{s+2}}^{(s+1)}\rangle\langle\tilde{\nu}_{z_{s+2}}^{(s+1)}|,\\
|\upsilon_{y_{s+2}}^{(s+2)}\rangle&=|\psi_{y_{s+2}}^{(s+2)}\rangle-\sum_{z_{s+2}=1}^{y_{s+2}}\langle\tilde{\nu}_{N_{s}+1-z_{s+2}}^{(s)}|\psi_{y_{s+2}}^{(s+2)}\rangle|\nu_{z_{s+2}}^{(s)}\rangle-\sum_{z_{s+2}=1}^{y_{s+2}}\langle\tilde{\nu}_{N_{s+1}+1-z_{s+2}}^{(s+1)}|\psi_{y_{s+2}}^{(s+2)}\rangle|\nu_{z_{s+2}}^{(s+1)}\rangle,
\end{split}
\end{equation}
where $\langle\tilde{\upsilon}_{y_{s+2}}^{(s+2)}|$ and $|\upsilon_{y_{s+2}}^{(s+2)}\rangle$ still obey the Jordan chain and $\langle\tilde{\upsilon}_{y_{s+2}}^{(s+2)}|\nu_{y_{s}}^{(s)}\rangle=\langle\tilde{\upsilon}_{y_{s+2}}^{(s+2)}|\nu_{y_{s+1}}^{(s+1)}\rangle=\langle\tilde{\nu}_{y_{s}}^{(s)}|\upsilon_{y_{s+2}}^{(s+2)}\rangle=\langle\tilde{\nu}_{y_{s+1}}^{(s+1)}|\upsilon_{y_{s+2}}^{(s+2)}\rangle=0$, and further redefine $\langle\tilde{\upsilon}_{y_{s+2}}^{(s+2)}|$ and $|\upsilon_{y_{s+2}}^{(s+2)}\rangle$ as $\langle\tilde{\nu}_{y_{s+2}}^{(s+2)}|$ and $|\nu_{y_{s+2}}^{(s+2)}\rangle$ by Eq.~(\ref{SEq-7}), with $C_{s+2}=\langle\tilde{\upsilon}_{N_{s+2}}^{(s+2)}|\upsilon_{1}^{(s+2)}\rangle=\langle\tilde{\psi}_{N_{s+2}}^{(s+2)}|\psi_{1}^{(s+2)}\rangle$.

(\romannumeral4) We repeat the procedures above and every time can generate a set of
\begin{equation}\label{SEq-11}
\begin{split}
\langle\tilde{\upsilon}_{y_{s+k}}^{(s+k)}|&=\langle\tilde{\psi}_{y_{s+k}}^{(s+k)}|-\sum_{h=0}^{k-1}\sum_{z_{s+k}=1}^{y_{s+k}}\langle\tilde{\psi}_{y_{s+k}}^{(s+k)}|\nu_{N_{s+h}+1-z_{s+k}}^{(s+h)}\rangle\langle\tilde{\nu}_{z_{s+k}}^{(s+h)}|,\\
|\upsilon_{y_{s+k}}^{(s+k)}\rangle&=|\psi_{y_{s+k}}^{(s+k)}\rangle-\sum_{h=0}^{k-1}\sum_{z_{s+k}=1}^{y_{s+k}}\langle\tilde{\nu}_{N_{s+h}+1-z_{s+k}}^{(s+h)}|\psi_{y_{s+k}}^{(s+k)}\rangle|\nu_{z_{s+k}}^{(s+h)}\rangle,
\end{split}
\end{equation}
which obey the Jordan chain and hold the biorthogonality $\langle\tilde{\upsilon}_{y_{s+k}}^{(s+k)}|\nu_{y_{s+h}}^{(s+h)}\rangle=\langle\tilde{\nu}_{y_{s+h}}^{(s+h)}|\upsilon_{y_{s+k}}^{(s+k)}\rangle=0$ ($h\in[0,k-1]$). We then employ Eq.~(\ref{SEq-7}) to redefine them as $\langle\tilde{\nu}_{y_{s+k}}^{(s+k)}|$ and $|\nu_{y_{s+k}}^{(s+k)}\rangle$, respectively, with $C_{s+k}=\langle\tilde{\upsilon}_{N_{s+k}}^{(s+k)}|\upsilon_{1}^{(s+k)}\rangle=\langle\tilde{\psi}_{N_{s+k}}^{(s+k)}|\psi_{1}^{(s+k)}\rangle$.

Finally, for each degenerate case, the orthonormalization among the eigenstates and generalized eigenstates of $E_{\mathrm{EP}}^{(s+k)\ast}$ and $E_{\mathrm{EP}}^{(s+k^{\prime})}$ is attained, and the PCR of $\hat{\mathcal{H}}$ now reads
\begin{equation}\label{SEq-12}
\begin{split}
\sum_{d=1}^{\mathbb{L}}\sum_{k=1}^{L_{d}}\sum_{y_{d|k}=1}^{N_{d|k}}|\nu_{y_{d|k}}^{(d|k)}\rangle\langle\tilde{\nu}_{N_{d|k}+1-y_{d|k}}^{(d|k)}|+\sum_{s=1}^{\mathbb{S}}\sum_{y_{s}=1}^{N_{s}}|\nu_{y_{s}}^{(s)}\rangle\langle\tilde{\nu}_{N_{s}+1-y_{s}}^{(s)}|+\sum_{r=1}^{\mathbb{R}}|\omega_{r}\rangle\langle\tilde{\omega}_{r}|=\hat{\mathbb{I}},
\end{split}  
\end{equation} 
where $\mathbb{S}=\mathbb{N}-\sum_{d=1}^{\mathbb{L}}L_{d}$ and $\mathbb{R}=M-\sum_{d=1}^{\mathbb{L}}\sum_{k=1}^{L_{d}}N_{d|k}-\sum_{s=1}^{\mathbb{S}}N_{s}$, with $d|k$ providing the label of the $k$th eigenenergy in the $d$th degenerate case, and $\langle\tilde{\nu}_{y_{d|k}}^{(d|k)}|\nu_{y_{d^{\prime}|k^{\prime}}^{\prime}}^{(d^{\prime}|k^{\prime})}\rangle=\delta_{d,d^{\prime}}\delta_{k,k^{\prime}}\delta_{y_{d|k},N_{d|k}+1-y_{d|k}^{\prime}}$, $\langle\tilde{\nu}_{y_{s}}^{(s)}|\nu_{y_{s^{\prime}}^{\prime}}^{(s^{\prime})}\rangle=\delta_{s,s^{\prime}}\delta_{y_{s},N_{s}+1-y_{s}^{\prime}}$, $\langle\tilde{\omega}_{r}|\omega_{r^{\prime}}\rangle=\delta_{r,r^{\prime}}$, and $\langle\tilde{\nu}_{y_{s}}^{(s)}|\nu_{y_{d|k}}^{(d|k)}\rangle=\langle\tilde{\nu}_{y_{d|k}}^{(d|k)}|\nu_{y_{s}}^{(s)}\rangle=\langle\tilde{\omega}_{r}|\nu_{y_{d|k}}^{(d|k)}\rangle=\langle\tilde{\nu}_{y_{d|k}}^{(d|k)}|\omega_{r}\rangle=\langle\tilde{\omega}_{r}|\nu_{y_{s}}^{(s)}\rangle=\langle\tilde{\nu}_{y_{s}}^{(s)}|\omega_{r}\rangle=0$. 

The corresponding EP dynamics is thus
\begin{equation}\label{SEq-13}
\begin{split}
|\Psi(t)\rangle=\sum_{d=1}^{\mathbb{L}}\sum_{k=1}^{L_{d}}\sum_{y_{d|k}=1}^{N_{d|k}}\sum_{p_{d|k}^{\prime}=0}^{N_{d|k}-y_{d|k}}\mathcal{C}_{y_{d|k},p_{d|k}^{\prime}}^{(d|k)}\left(t\right)t^{p_{d|k}^{\prime}}|\nu_{y_{d|k}}^{(d|k)}\rangle+\sum_{s=1}^{\mathbb{S}}\sum_{y_{s}=1}^{N_{s}}\sum_{p_{s}^{\prime}=0}^{N_{s}-y_{s}}\mathcal{C}_{y_{s},p_{s}^{\prime}}^{(s)}\left(t\right)t^{p_{s}^{\prime}}|\nu_{y_{s}}^{(s)}\rangle+\sum_{r=1}^{\mathbb{R}}\mathcal{C}_{r}\left(t\right)|\omega_{r}\rangle.
\end{split}
\end{equation}
Here, $\mathcal{C}_{y_{d|k},p_{d|k}^{\prime}}^{(d|k)}(t)=(-i)^{p_{d|k}^{\prime}}\zeta_{y_{d|k}+p_{d|k}^{\prime}}^{(d|k)}e^{-iE_{\mathrm{EP}}^{(d|k)}t}/p_{d|k}^{\prime}!$, with $E_{\mathrm{EP}}^{(d|k)}$ the eigenenergy for the $d$th degenerate case and $\zeta_{y_{d|k}}^{(d|k)}=\langle\tilde{\nu}_{N_{d|k}+1-y_{d|k}}^{(d|k)}|\Psi(0)\rangle$, and the definitions of $\mathcal{C}_{y_{s},p_{s}^{\prime}}^{(s)}(t)$ and $\mathcal{C}_{r}(t)$ remain unchanged, as given in the main text.

It should be emphasized that we need to be careful when the algebraic multiplicities of some $E_{\mathrm{EP}}^{(d|k)}$ in each degenerate case are equal, since the left partner of a right (generalized) eigenstate may not reside in the present subspace. For instance, consider $N_{d|k}=N_{d|k+1}$, the left partner of $|\psi_{1}^{(d|k)}\rangle$ may be $\langle\tilde{\psi}_{N_{d|k+1}}^{(d|k+1)}|$ rather than $\langle\tilde{\psi}_{N_{d|k}}^{(d|k)}|$, due to the fact that the biorthogonality $\langle\tilde{\psi}_{N_{d|k+1}}^{(d|k+1)}|\psi_{1}^{(d|k)}\rangle$ is also unknown. If this is true, the left partner of $|\psi_{1}^{(d|k+1)}\rangle$ must be $\langle\tilde{\psi}_{N_{d|k}}^{(d|k)}|$. Similarly, the left partner of $|\psi_{2}^{(d|k)}\rangle$ may be $\langle\tilde{\psi}_{N_{d|k}-1}^{(d|k)}|$ or $\langle\tilde{\psi}_{N_{d|k+1}-1}^{(d|k+1)}|$, due to the fact that $\langle\tilde{\psi}_{N_{d|k}}^{(d|k)}|\psi_{2}^{(d|k)}\rangle$, $\langle\tilde{\psi}_{N_{d|k+1}}^{(d|k+1)}|\psi_{2}^{(d|k)}\rangle$, $\langle\tilde{\psi}_{N_{d|k}-1}^{(d|k)}|\psi_{2}^{(d|k)}\rangle$, and $\langle\tilde{\psi}_{N_{d|k+1}-1}^{(d|k+1)}|\psi_{2}^{(d|k)}\rangle$ are all unknown and $\langle\tilde{\psi}_{N_{d|k}}^{(d|k)}|$ and $\langle\tilde{\psi}_{N_{d|k+1}}^{(d|k+1)}|$ have been matched with $|\psi_{1}^{(d|k+1)}\rangle$ and $|\psi_{1}^{(d|k)}\rangle$, respectively, so that when the left partner of $|\psi_{2}^{(d|k)}\rangle$ is determined be one of them, the left partner of $|\psi_{2}^{(d|k+1)}\rangle$ must be the other. Hence, in this scenario, we first need to find out the left partner of each right (generalized) eigenstate correctly based on the actual situation, then, we redefine them according to the orthonormalization procedure to establish the corresponding PCR. While these redefined generalized eigenstates may no longer obey the Jordan chain, the eigenstates of these degenerate eigenenergies with the same algebraic multiplicity $N_{d|k}$ still hold the polynomial growth with degree $N_{d|k}-1$ over time. Refer to the case of two degenerate 2nd-order EPs in the NH diamond ring.  

\subsection{Pseudo-completeness relations for the presence of more degeneracy}	
If some of $E_{r}$ are also degenerate with $E_{\mathrm{EP}}^{(d|k)}$, and there are $\mathbb{G}$ cases where each has $G_{d_{c}}$ ($c\in[1,\mathbb{G}]$) degenerate eigenenergies, with $d_{c}$ indicating the concrete value of $d$ in the $c$th case, all $|\phi_{u}^{(d_{c})}\rangle$ ($u\in[1,G_{d_{c}}]$) are also the eigenstates of $\hat{\mathcal{A}}_{\mathbb{N}}$ with a $(\sum_{k=1}^{L_{d_{c}}}N_{d_{c}|k}+G_{d_{c}})$-fold degenerate eigenenergy $\mathbb{E}_{\mathrm{DP}}^{(d_{c}|k)}$, and all $|\tilde{\phi}_{u}^{(d_{c})}\rangle$ are also the eigenstates of $\hat{\mathcal{A}}_{\mathbb{N}}^{\dagger}$ with a $(\sum_{k=1}^{L_{d_{c}}}N_{d_{c}|k}+G_{d_{c}})$-fold degenerate eigenenergy $\mathbb{E}_{\mathrm{DP}}^{(d_{c}|k)\ast}$. Therefore, the biorthogonality between $\langle\tilde{\nu}_{y_{d_{c}|k}}^{(d_{c}|k)}|$ and $|\phi_{u}^{(d_{c})}\rangle$, between $\langle\tilde{\phi}_{u}^{(d_{c})}|$ and $|\nu_{y_{d_{c}|k}}^{(d_{c}|k)}\rangle$, and between $\langle\tilde{\phi}_{u}^{(d_{c})}|$ and $|\phi_{u^{\prime}}^{(d_{c})}\rangle$ ($u\neq u^{\prime}$) should be also guaranteed. To this end, we first rewrite $\langle\tilde{\phi}_{u}^{(d_{c})}|$ and $|\phi_{u}^{(d_{c})}\rangle$ as
\begin{equation}\label{SEq-14}
\begin{split}
\langle\tilde{\varphi}_{u}^{(d_{c})}|&=\langle\tilde{\phi}_{u}^{(d_{c})}|-\sum_{k=1}^{L_{d_{c}}}\langle\tilde{\phi}_{u}^{(d_{c})}|\nu_{N_{d_{c}|k}}^{(d_{c}|k)}\rangle\langle\tilde{\nu}_{1}^{(d_{c}|k)}|-\sum_{o=1}^{u-1}\langle\tilde{\phi}_{u}^{(d_{c})}|\omega_{o}^{(d_{c})}\rangle\langle\tilde{\omega}_{o}^{(d_{c})}|,\\
|\varphi_{u}^{(d_{c})}\rangle&=|\phi_{u}^{(d_{c})}\rangle-\sum_{k=1}^{L_{d_{c}}}\langle\tilde{\nu}_{N_{d_{c}|k}}^{(d_{c}|k)}|\phi_{u}^{(d_{c})}\rangle|\nu_{1}^{(d_{c}|k)}\rangle-\sum_{o=1}^{u-1}\langle\tilde{\omega}_{o}^{(d_{c})}|\phi_{u}^{(d_{c})}\rangle|\omega_{o}^{(d_{c})}\rangle.
\end{split}
\end{equation}
Next, we normalize $\langle\tilde{\varphi}_{u}^{(d_{c})}|$ and $|\varphi_{u}^{(d_{c})}\rangle$ as $\langle\tilde{\omega}_{u}^{(d_{c})}|$ and $|\omega_{u}^{(d_{c})}\rangle$, respectively, by Eq.~(\ref{SEq-7}). Finally, equation~(\ref{SEq-12}) becomes
\begin{equation}\label{SEq-15}
\begin{split}
\sum_{d=1}^{\mathbb{L}}\sum_{k=1}^{L_{d}}\sum_{y_{d|k}=1}^{N_{d|k}}|\nu_{y_{d|k}}^{(d|k)}\rangle\langle\tilde{\nu}_{N_{d|k}+1-y_{d|k}}^{(d|k)}|+\sum_{s=1}^{\mathbb{S}}\sum_{y_{s}=1}^{N_{s}}|\nu_{y_{s}}^{(s)}\rangle\langle\tilde{\nu}_{N_{s}+1-y_{s}}^{(s)}|+\sum_{c=1}^{\mathbb{G}}\sum_{u=1}^{G_{d_{c}}}|\omega_{u}^{(d_{c})}\rangle\langle\tilde{\omega}_{u}^{(d_{c})}|+\sum_{r=1}^{\mathbb{F}}|\omega_{r}\rangle\langle\tilde{\omega}_{r}|=\hat{\mathbb{I}},
\end{split}  
\end{equation}
with $\mathbb{F}=\mathbb{R}-\sum_{c=1}^{\mathbb{G}}G_{d_{c}}$. Note that for the case of a single EP, equation~(\ref{SEq-15}) reduces to
\begin{equation}\label{SEq-16}
\begin{split}
&\sum_{y_{1}=1}^{N_{1}}|\nu_{y_{1}}\rangle\langle\tilde{\nu}_{N_{1}+1-y_{1}}|+\sum_{u=1}^{G_{1}}|\omega_{u}^{(1)}\rangle\langle\tilde{\omega}_{u}^{(1)}|+\sum_{r=1}^{\mathbb{J}}|\omega_{r}\rangle\langle\tilde{\omega}_{r}|=\hat{\mathbb{I}},
\end{split}  
\end{equation}  
with $\mathbb{J}=M-N_{1}-G_{1}$, corresponding to the cases of the single 2nd-order EP in the NH stub ribbon and the single 3rd-order EP in the NH diamond ring. Additionally, except for the $\mathbb{G}$ cases, when there are $\mathbb{K}$ extra cases where in each case $K_{w}$ ($w\in[1,\mathbb{K}]$) eigenenergies $E_{r}$ are degenerate with each other, we should guarantee the biorthogonality between $\langle\tilde{\phi}_{v}^{(w)}|$ and $|\phi_{v^{\prime}}^{(w)}\rangle$ ($v,v^{\prime}\in[1,K_{w}]~\mathrm{and}~v\neq v^{\prime}$). Consequently, we rewrite $\langle\tilde{\phi}_{v}^{(w)}|$ and $|\phi_{v}^{(w)}\rangle$ as
\begin{equation}\label{SEq-17}
\begin{split}
\langle\tilde{\varphi}_{v}^{(w)}|&=\langle\tilde{\phi}_{v}^{(w)}|-\sum_{o=1}^{v-1}\langle\tilde{\phi}_{v}^{(w)}|\omega_{o}^{(w)}\rangle\langle\tilde{\omega}_{o}^{(w)}|,\\
|\varphi_{v}^{(w)}\rangle&=|\phi_{v}^{(w)}\rangle-\sum_{o=1}^{v-1}\langle\tilde{\omega}_{o}^{(w)}|\phi_{v}^{(w)}\rangle|\omega_{o}^{(w)}\rangle,
\end{split}
\end{equation} 
and further normalize them as $\langle\tilde{\omega}_{v}^{(w)}|$ and $|\omega_{v}^{(w)}\rangle$ by Eq.~(\ref{SEq-7}). The resulting PCR of $\hat{\mathcal{H}}$ is given by   
\begin{equation}\label{SEq-18}
\begin{split}
&\sum_{d=1}^{\mathbb{L}}\sum_{k=1}^{L_{d}}\sum_{y_{d|k}=1}^{N_{d|k}}|\nu_{y_{d|k}}^{(d|k)}\rangle\langle\tilde{\nu}_{N_{d|k}+1-y_{d|k}}^{(d|k)}|+\sum_{s=1}^{\mathbb{S}}\sum_{y_{s}=1}^{N_{s}}|\nu_{y_{s}}^{(s)}\rangle\langle\tilde{\nu}_{N_{s}+1-y_{s}}^{(s)}|+\sum_{c=1}^{\mathbb{G}}\sum_{u=1}^{G_{d_{c}}}|\omega_{u}^{(d_{c})}\rangle\langle\tilde{\omega}_{u}^{(d_{c})}|+\sum_{w=1}^{\mathbb{K}}\sum_{v=1}^{K_{w}}|\omega_{v}^{(w)}\rangle\langle\tilde{\omega}_{v}^{(w)}|\\
&+\sum_{r=1}^{\mathbb{P}}|\omega_{r}\rangle\langle\tilde{\omega}_{r}|=\hat{\mathbb{I}},
\end{split}  
\end{equation}
with $\mathbb{P}=\mathbb{F}-\sum_{w=1}^{\mathbb{K}}K_{w}$, and we can further obtain the EP dynamics as
\begin{equation}\label{SEq-19}
\begin{split}
|\Psi(t)\rangle=&\sum_{d=1}^{\mathbb{L}}\sum_{k=1}^{L_{d}}\sum_{y_{d|k}=1}^{N_{d|k}}\sum_{p_{d|k}^{\prime}=0}^{N_{d|k}-y_{d|k}}\mathcal{C}_{y_{d|k},p_{d|k}^{\prime}}^{(d|k)}\left(t\right)t^{p_{d|k}^{\prime}}|\nu_{y_{d|k}}^{(d|k)}\rangle+\sum_{s=1}^{\mathbb{S}}\sum_{y_{s}=1}^{N_{s}}\sum_{p_{s}^{\prime}=0}^{N_{s}-y_{s}}\mathcal{C}_{y_{s},p_{s}^{\prime}}^{(s)}\left(t\right)t^{p_{s}^{\prime}}|\nu_{y_{s}}^{(s)}\rangle+\sum_{c=1}^{\mathbb{G}}\sum_{u=1}^{G_{d_{c}}}\mathcal{C}_{u}^{(d_{c})}\left(t\right)|\omega_{u}^{(d_{c})}\rangle\\
&+\sum_{w=1}^{\mathbb{K}}\sum_{v=1}^{K_{w}}\mathcal{C}_{v}^{(w)}\left(t\right)|\omega_{v}^{(w)}\rangle+\sum_{r=1}^{\mathbb{R}}\mathcal{C}_{r}\left(t\right)|\omega_{r}\rangle.
\end{split}
\end{equation}
Here, $\mathcal{C}_{u}^{(d_{c})}(t)=\iota_{u}^{(d_{c})}e^{-iE_{\mathrm{EP}}^{(d_{c}|k)}t}$, with $\iota_{u}^{(d_{c})}=\langle\tilde{\omega}_{u}^{(d_{c})}|\Psi(0)\rangle$, and $\mathcal{C}_{v}^{(w)}(t)=\iota_{v}^{(w)}e^{-iE^{(w)}t}$, with $E^{(w)}$ the eigenenergy for the $w$th extra degenerate case and $\iota_{v}^{(w)}=\langle\tilde{\omega}_{v}^{(w)}|\Psi(0)\rangle$. 	

\section{Flat band and compact localized state of Hermitian stub ribbon}

\subsection{Flat band}
\begin{figure}\centering
	\includegraphics[width=0.8\linewidth]{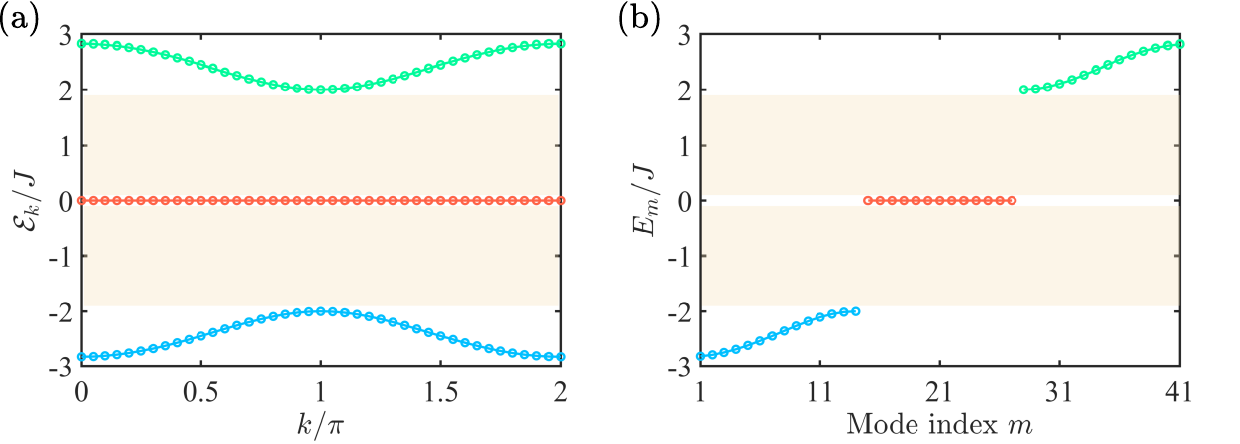}
	\caption{(a) Three dispersion relations for the Hermitian stub ribbon, where sublattice $C$ is present in the last unit cell and $J_{n}^{u}=J_{n}^{d}=\Lambda=2J$. A dispersionless zero-energy flat band can be found. (b) Spectrum of a finite Hermitian stub ribbon with $\Lambda=2J$ and $N=14$ illustrated in the main text. One can observe that the plateau of the zero-energy flat band with a $13$-fold degeneracy is reproduced in the middle, causing the occurrence of a high-order DP.}\label{figS1}
\end{figure}

From the perspective of dispersion, the Hermitian stub ribbon supports a dispersionless flat band lying at the zero energy. To see this clearly, it is necessary for guaranteeing the translation invariance of the model to assign $J_{n}^{u}=J_{n}^{d}=\Lambda$ for all $n\in[1,N]$ and to temporarily replenish the site $C_{N}$ in the last unit cell simultaneously. Further, under the periodic boundary condition, we can safely execute the Fourier transformation, $\hat{o}_{n}=1/\sqrt{N}\sum_{k}e^{ikn}\hat{o}_{k}$ ($\hat{o}=\hat{a},\hat{b}, \hat{c}$), with $k\in[0,2\pi]$ the wave vector of the lattice in the first Brillouin zone, to inspect its dispersion relation in the momentum space, so that the resulting Hamiltonian of the system can be written in the form of 
\begin{equation}\label{Eq1}
\hat{\mathcal{H}}=\sum_{k}\hat{\Phi}_{k}^{\dagger}\hat{h}\left(k\right)\hat{\Phi}_{k},
\end{equation}
where $\hat{\Phi}_{k}^{\dagger}=(\hat{a}_{k}^{\dagger},\hat{b}_{k}^{\dagger},\hat{c}_{k}^{\dagger})$ and 
\begin{equation}\label{Eq2}
\hat{h}\left(k\right)=\begin{pmatrix}
~0~&~\Lambda~&~0\\
~\Lambda~&~0~&~J\left(1+e^{-ik}\right)\\
~0~&~J\left(1+e^{ik}\right)~&~0\\
\end{pmatrix}.
\end{equation}
By diagonalizing Eq.~(\ref{Eq2}), three dispersion relations can be obtained as $0$ and $\pm\sqrt{\Lambda^{2}+2J^{2}(1+\cos k)}$, respectively, whereupon the Hermitian counterpart features one zero-energy flat band independent of $k$. Moreover, it is worth mentioning that $\hat{h}(k)$ respects the chiral symmetry (CS), $\hat{V}\hat{h}(k)\hat{V}^{-1}=-\hat{h}(k)$, with the unitary chiral operator 
\begin{equation}\label{Eq3}
\hat{V}=\begin{pmatrix}
~1~&~0~&~0~\\
~0~&~-1~&~0~\\
~0~&~0~&~1~\\
\end{pmatrix},
\end{equation} 
which implies that these bands will emerge in pairs, $\lbrace \mathcal{E}_{k},-\mathcal{E}_{k}\rbrace$, bringing about a symmetric spectrum and a dispersionless zero-energy flat band protected by the CS. As a consequence, there is one eigenstate $|\varphi_{k}\rangle$ with eigenenergy $\mathcal{E}_{k}$ and the other partner $\hat{V}|\varphi_{k}\rangle$ with eigenenergy $-\mathcal{E}_{k}$. Besides, the minimum gap between two adjacent bands takes $\Lambda$ at $k=\pi$. As an example, we plot in Fig.~\ref{figS1}(a) these three dispersion relations when $\Lambda=2J$.

On the other hand, figure \ref{figS1}(b) shows the spectrum of a finite Hermitian counterpart with $\Lambda=2J$ and $N=14$ illustrated in the main text by imposing the open boundary condition. One can observe that the upper and lower bulk bands are also symmetric, and the spectrum reproduces the plateau of the zero-energy flat band in the middle, accompanied with a $13$-fold degeneracy and further the occurrence of a high-order diabolic point (DP), which is intimately attributed to the preservation of the CS, with the real-space unitary chiral operator $\hat{\Gamma}=(\bigoplus_{j=1}^{13}\hat{V})\bigoplus\hat{\sigma_{z}}$, where $\hat{V}$ is given in Eq.~(\ref{Eq3}) and $\hat{\sigma_{z}}$ stands for the Pauli-$z$ operator. For a general finite Hermitian case of $J_{n}^{u}=J_{n}^{d}=\Lambda_{n}$ with $n\in[1,N]$, it turns out that the CS is still respected, with the relevant unitary chiral operator
\begin{equation}\label{Eq4}
\hat{\Gamma}=\left[\bigoplus_{j=1}^{N-1}\hat{V}\right]\bigoplus\hat{\sigma_{z}},
\end{equation}
and the symmetric spectrum is characterized by one $(N-1)$-fold degenerate zero-energy flat band responsible for a DP of order $N-1$.

\subsection{Compact localized state}
The eigenstates of the $(N-1)$-fold degenerate zero-energy flat band are made up of the linear-independent compact localized states, each of which occupies only several sites inside two successive unit cells due to the destructive interference between probability amplitudes. To solve these compact localized states analytically, we commence with the energy eigenequation of a general finite Hermitian stub ribbon, $\hat{H}|\psi_{m}\rangle=E_{m}|\psi_{m}\rangle$, with the mode index $m\in[1,3N-1]$, and the $m$th eigenstate $|\psi_{m}\rangle$ is spanned as $|\psi_{m}\rangle=\sum_{n}\alpha_{m,n}\hat{a}_{n}^{\dagger}|\mathbf{0}\rangle+\beta_{m,n}\hat{b}_{n}^{\dagger}|\mathbf{0}\rangle+\gamma_{m,n}\hat{c}_{n}^{\dagger}|\mathbf{0}\rangle$, where $|\mathbf{0}\rangle$ is the vacuum state and $\alpha_{m,n}$, $\beta_{m,n}$, and $\gamma_{m,n}$ are the probability amplitudes on the sites $A_{n}$, $B_{n}$, and $C_{n}$, respectively, leading further to the following set of equations for these probability amplitudes
\begin{equation}\label{Eq5}
\begin{split}
E_{m}\alpha_{m,n}&=\Lambda_{n}\beta_{m,n},\\
E_{m}\beta_{m,n}&=\Lambda_{n}\alpha_{m,n}+J\left(\gamma_{m,n-1}+\gamma_{m,n}\right),\\
E_{m}\gamma_{m,n}&=J\left(\beta_{m,n}+\beta_{m,n+1}\right),
\end{split}
\end{equation} 
with the boundary conditions $\gamma_{m,0}=\gamma_{m,N}=0$. For the sake of simplification, we below express $|A_{n}\rangle=\hat{a}_{n}^{\dagger}|\mathbf{0}\rangle$, $|B_{n}\rangle=\hat{b}_{n}^{\dagger}|\mathbf{0}\rangle$, and $|C_{n}\rangle=\hat{c}_{n}^{\dagger}|\mathbf{0}\rangle$, respectively. For these $N-1$ zero-energy compact localized states $|\xi_{l}\rangle$, with $l\in[1,N-1]$, it is found that $\beta_{l,n}=0$ for all $n\in[1,N]$, $\alpha_{l,n}$ and $\gamma_{l,n}$ can destructively interfere at the site $B_{n}$, $\Lambda_{n}\alpha_{l,n}+J\gamma_{l,n}=0$, meanwhile, $\alpha_{l,n+1}$ and $\gamma_{l,n}$ can also destructively interfere at the site $B_{n+1}$, $\Lambda_{n+1}\alpha_{l,n+1}+J\gamma_{l,n}=0$, and the remainder vanish, which means that the nonvanishing probability amplitudes are distributed only on the sites $A_{n}$, $C_{n}$, and $A_{n+1}$ inside two adjacent unit cells. Consequently, the $l$th unnormalized zero-energy compact localized states can be canonically formulated as
\begin{equation}\label{Eq6}
\left|\xi_{l}\right\rangle=\frac{1}{\Lambda_{l}}\left|A_{l}\right\rangle+\frac{1}{\Lambda_{l+1}}\left|A_{l+1}\right\rangle-\frac{1}{J}\left|C_{l}\right\rangle.
\end{equation}

\section{Longer-time spontaneous splitting dynamics of two ratios with different initial states in non-Hermitian stub ribbon}

\begin{figure}\centering
	\includegraphics[width=\linewidth]{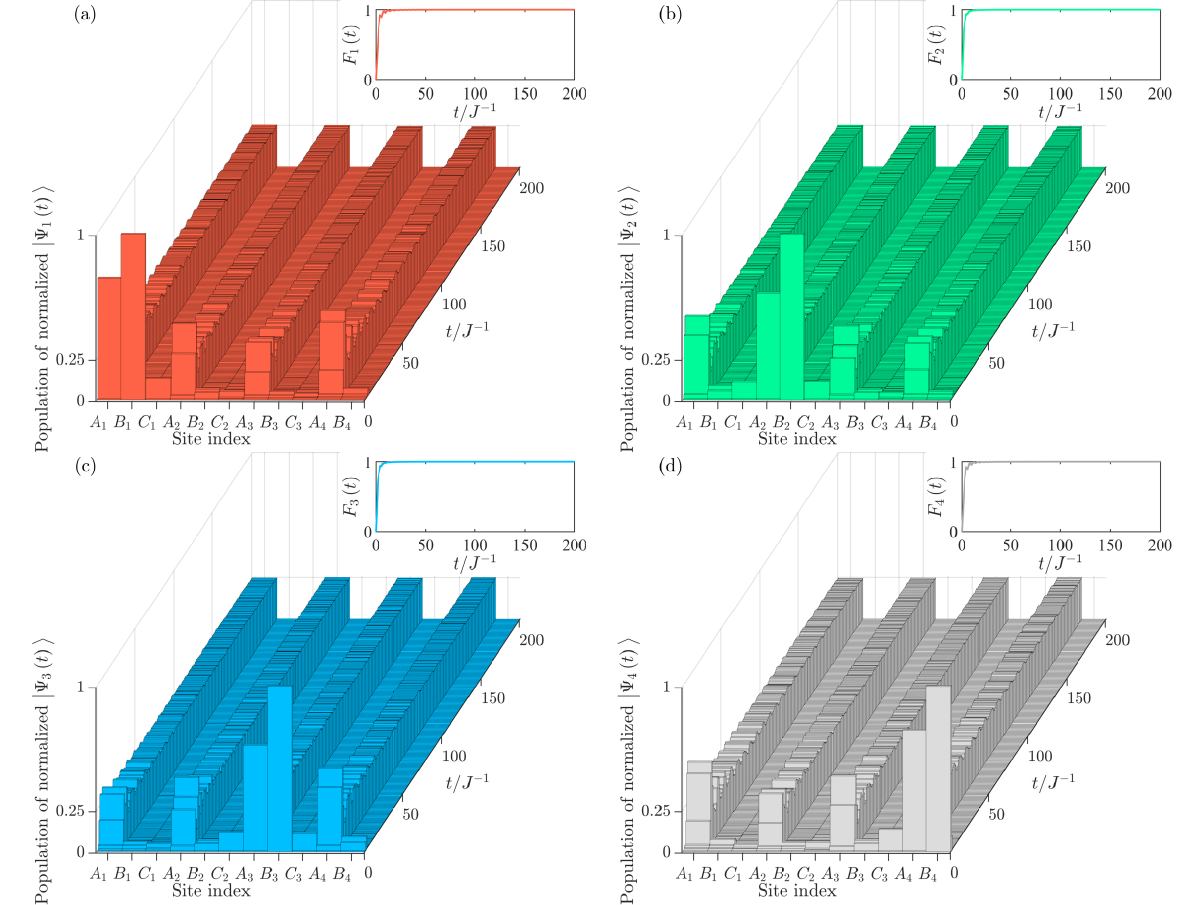}
	\caption{Population dynamics of normalized $|\Psi_{n}(t)\rangle$ when initial states are chosen as (a) $|\Psi_{1}(0)\rangle=|B_{1}\rangle$, (b) $|\Psi_{2}(0)\rangle=|B_{2}\rangle$, (c) $|\Psi_{3}(0)\rangle=|B_{3}\rangle$, and (d) $|\Psi_{4}(0)\rangle=|B_{4}\rangle$, respectively, all of which show a $4$-port spontaneous splitting process with ratio $25\colon25\colon25\colon25$. The inset in each panel shows the corresponding fidelity $F_{n}(t)$. The parameters of the model are $N=4$, $\lambda=0$, and $\lbrace J_{n}^{u}=2J\rbrace$.}\label{figS2}
\end{figure}

\begin{figure}\centering
	\includegraphics[width=\linewidth]{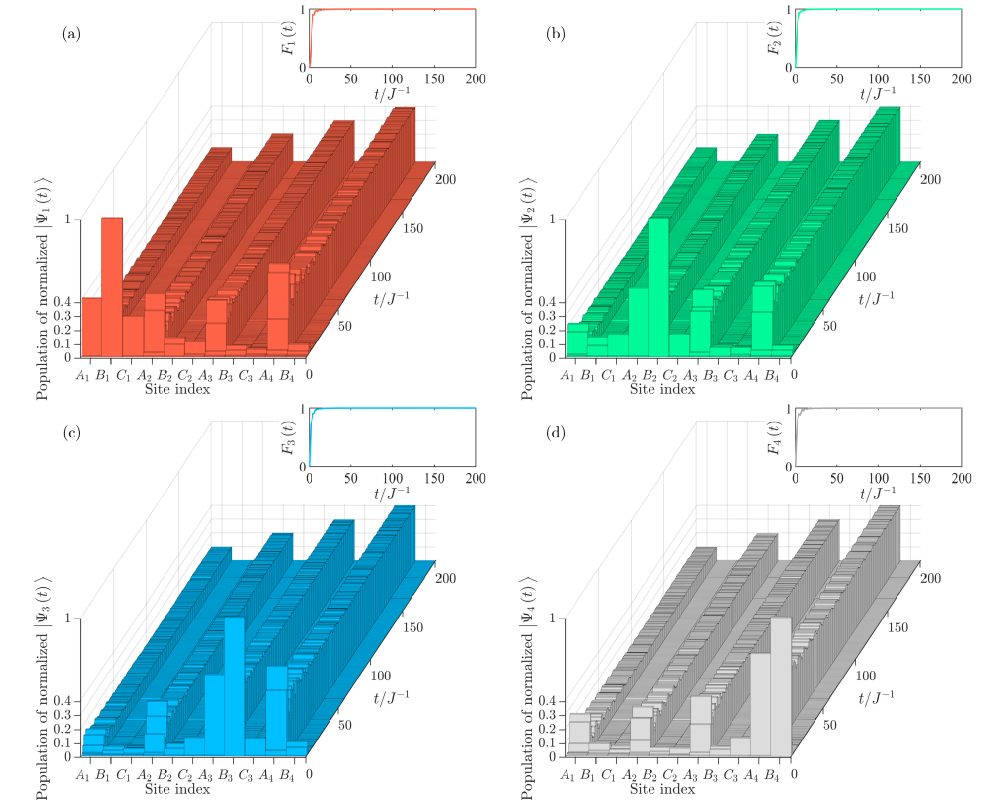}
	\caption{(a)-(d) Same as Figs.~\ref{figS2}(a)-\ref{figS2}(d) except for $\lbrace J_{n}^{u}=\sqrt{n}J\rbrace$, showing a $10\colon20\colon30\colon40$ ratio.}\label{figS3}
\end{figure}

For $N=4$, we plot in Figs.~\ref{figS2}(a)-\ref{figS2}(d) the EP dynamics of the NH stub ribbon with $\lbrace J_{n}^{u}=2J\rbrace$ in longer-term evolution at $\lambda=0$ when the initial states are chosen as $\lbrace|\Psi_{n}(0)\rangle=|B_{n}\rangle\rbrace$. One can observe from the population of each normalized $|\Psi_{n}(t)\rangle$ that regardless of which site $B_{n}$ the initial excitation is injected into, it is spontaneously distributed only onto all the sublattices $A$ with an almost equal probability forever, where the presence of $|\omega_{f}^{\mathrm{EP}}\rangle$ gives rise to the marginal oscillations that become negligible for $t\rightarrow\infty$, so that a spontaneous $25\colon25\colon25\colon25$ 4-port beam splitter is fabricated successfully. Figures~\ref{figS3}(a)-\ref{figS3}(d) are the same as Figs.~\ref{figS2}(a)-\ref{figS2}(d) with the exception of $\lbrace J_{n}^{u}=\sqrt{n}J\rbrace$. Visibly, an arithmetic splitting ratio of $10\colon20\colon30\colon40$ can be achieved. Furthermore, the fidelity $F_{n}(t)=|\langle\nu^{\mathrm{EP}}_{\mathrm{nor}}|\Psi_{n}(t)\rangle|^{2}$ is also introduced to quantitatively measure the degree to which the normalized $|\Psi_{n}(t)\rangle$ and $|\nu^{\mathrm{EP}}_{\mathrm{nor}}\rangle$ with $|\nu^{\mathrm{EP}}_{\mathrm{nor}}\rangle=|\nu^{\mathrm{EP}}\rangle/\sqrt{\langle\nu^{\mathrm{EP}}|\nu^{\mathrm{EP}}\rangle}$ overlap within evolution, as shown in the inset of each panel, where $F_{n}(t)\rightarrow1$ for $t\rightarrow\infty$ means that the spontaneous conversion of $|\Psi_{n}(t)\rangle$ to $|\nu^{\mathrm{EP}}\rangle$ is accomplished and further characterize the realization of the beam splitter. Moreover, we can also find that the efficiency for the state conversion is exceedingly pronounced, which suggests that the spontaneous splitting process occurs rapidly.

\section{Eigenstates and generalized eigenstates of system at different exceptional points in non-Hermitian diamond ring}

For real $\epsilon$ or imaginary $\epsilon$ ($\epsilon\neq\pm i$) and $\kappa=1$, the unnormalized eigenstate $|\psi_{1}\rangle$ and two generalized eigenstates $|\psi_{2}\rangle$ and $|\psi_{3}\rangle$ of the 3rd-order EP and the eigenstate $|\phi\rangle$ of the extra degenerate zero eigenenergy are given in Table~\ref{tab1}, where their matching left partners are $\langle\psi_{3}^{\ast}|$, $\langle\psi_{2}^{\ast}|$, $\langle\psi_{1}^{\ast}|$, and $\langle\phi^{\ast}|$, respectively, with ``$\ast$'' the complex conjugate operation, due to $\hat{H}^{\mathrm{T}}=\hat{H}$. 

We can find that all the biorthogonality works except that $\langle\psi_{3}^{\ast}|\psi_{3}\rangle$ now, so that in Table~\ref{tab1} we rewrite $|\psi_{3}\rangle$ as $|\varphi_{3}\rangle$, which is obtained from previous universal theory, and further redefine $\langle\tilde{\nu}_{3}|=\langle\psi_{3}^{\ast}|/\sqrt{-2i(1+\epsilon^{2})}$, $|\nu_{1}\rangle=|\psi_{1}\rangle/\sqrt{-2i(1+\epsilon^{2})}$, $\langle\tilde{\nu}_{2}|=\langle\psi_{2}^{\ast}|/\sqrt{-2i(1+\epsilon^{2})}$, $|\nu_{2}\rangle=|\psi_{2}\rangle/\sqrt{-2i(1+\epsilon^{2})}$,  $|\nu_{3}\rangle=|\varphi_{3}\rangle/\sqrt{-2i(1+\epsilon^{2})}$, $\langle\tilde{\nu}_{1}|=\langle\psi_{1}^{\ast}|/\sqrt{-2i(1+\epsilon^{2})}$, $\langle\tilde{\omega}|=\langle\phi^{\ast}|/\sqrt{1+\epsilon^{2}}$ and $|\omega\rangle=|\phi\rangle/\sqrt{1+\epsilon^{2}}$. The PCR is thus attained as 
\begin{equation}\label{SEq-3-1}
\begin{split}
\sum_{n=1}^{3}|\nu_{n}\rangle\langle\tilde{\nu}_{4-n}|+|\omega\rangle\langle\tilde{\omega}|=\hat{\mathbb{I}}.
\end{split}
\end{equation}
We also show in Table~\ref{tab2} the case of real $\epsilon$ and $\kappa=-1$, and the PCR becomes $\sum_{n=1}^{3}|\nu_{n}^{\ast}\rangle\langle\tilde{\nu}_{4-n}^{\ast}|+|\omega^{\ast}\rangle\langle\tilde{\omega}^{\ast}|=\hat{\mathbb{I}}$ due to $\hat{H}(\kappa=-1)=\hat{H}^{\dagger}(\kappa=1)$.

\begin{table}[htpb]
	\centering
	\caption{Distributions of eigenstates $|\psi_{1}\rangle$ and $|\phi\rangle$, generalized eigenstates $|\psi_{2}\rangle$ and $|\psi_{3}\rangle$, and redefined biorthogonal generalized eigenstate $|\varphi_{3}\rangle$ of system at the 3rd-order EP for real or imaginary $\epsilon$ ($\epsilon\neq\pm i$) and $\kappa=1$. Their matching left partners and the corresponding biorthogonal norms are also listed, respectively.}
	\setlength{\tabcolsep}{5.9mm}{
		\begin{tabular}{c c c c c c c}\hline\hline
			& $|A\rangle$ & $|B\rangle$ & $|C\rangle$ & $|D\rangle$ & Left partner & Biorthogonal norm \\ \hline
			$|\psi_{1}\rangle$ & 0 & $2(1+\epsilon^{2})$ & 0 & $-2i(1+\epsilon^{2})$ & $\langle\psi_{3}^{\ast}|$ & $\sqrt{-2i(1+\epsilon^{2})}$ \\ 
			$|\psi_{2}\rangle$ & $1-i$ & 0 & $\epsilon(1-i)$ & 0 & $\langle\psi_{2}^{\ast}|$ & $\sqrt{-2i(1+\epsilon^{2})}$ \\ 
			$|\psi_{3}\rangle$ & 0 & 0 & 0 & 1 & $\backslash$ & $\backslash$ \\ 
			$|\varphi_{3}\rangle$ & 0 & $-i$ & 0 & 0 & $\langle\psi_{1}^{\ast}|$ & $\sqrt{-2i(1+\epsilon^{2})}$ \\ 
			$|\phi\rangle$ & $-\epsilon$ & 0 & $1$ & 0 & $\langle\phi^{\ast}|$ & $\sqrt{1+\epsilon^{2}}$ \\ \hline\hline
		\end{tabular}
	}
	\label{tab1}
\end{table}

\begin{table}[htpb]
	\centering
	\caption{Same as Table~\ref{tab1} except for $\kappa=-1$.}
	\setlength{\tabcolsep}{6.07mm}{
		\begin{tabular}{c c c c c c c}\hline\hline
			& $|A\rangle$ & $|B\rangle$ & $|C\rangle$ & $|D\rangle$ & Left partner & Biorthogonal norm \\ \hline
			$|\psi_{1}\rangle$ & 0 & $2(1+\epsilon^{2})$ & 0 & $2i(1+\epsilon^{2})$ & $\langle\psi_{3}^{\ast}|$ & $\sqrt{2i(1+\epsilon^{2})}$ \\ 
			$|\psi_{2}\rangle$ & $1+i$ & 0 & $\epsilon(1+i)$ & 0 & $\langle\psi_{2}^{\ast}|$ & $\sqrt{2i(1+\epsilon^{2})}$ \\ 
			$|\psi_{3}\rangle$ & 0 & 0 & 0 & 1 & $\backslash$ & $\backslash$ \\ 
			$|\varphi_{3}\rangle$ & 0 & $i$ & 0 & 0 & $\langle\psi_{1}^{\ast}|$ & $\sqrt{2i(1+\epsilon^{2})}$ \\ 
			$|\phi\rangle$ & $-\epsilon$ & 0 & $1$ & 0 & $\langle\phi^{\ast}|$ & $\sqrt{1+\epsilon^{2}}$ \\ \hline\hline
		\end{tabular}
	}
	\label{tab2}
\end{table}

Similarly, Tables~\ref{tab3} and~\ref{tab4} give $|\psi_{1}\rangle$, $|\psi_{2}\rangle$, $|\psi_{3}\rangle$, and $|\phi\rangle$ for $\epsilon=\pm i$ and $\kappa\neq\pm1$, respectively, with their matching left partners $\langle\psi_{3}^{\ast}|$, $\langle\psi_{2}^{\ast}|$, $\langle\psi_{1}^{\ast}|$, and $\langle\phi^{\ast}|$. We still need to rewrite $|\psi_{3}\rangle$ as $|\varphi_{3}\rangle$ and to further redefine $\langle\tilde{\nu}_{3}|=\langle\psi_{3}^{\ast}|/\sqrt{2(\kappa^{2}-1)}$, $|\nu_{1}\rangle=|\psi_{1}\rangle/\sqrt{2(\kappa^{2}-1)}$,  $\langle\tilde{\nu}_{2}|=\langle\psi_{2}^{\ast}|/\sqrt{2(\kappa^{2}-1)}$, $|\nu_{2}\rangle=|\psi_{2}\rangle/\sqrt{2(\kappa^{2}-1)}$,  $\langle\tilde{\nu}_{1}|=\langle\psi_{1}^{\ast}|/\sqrt{2(\kappa^{2}-1)}$, $|\nu_{3}\rangle=|\varphi_{3}\rangle/\sqrt{2(\kappa^{2}-1)}$,  $\langle\tilde{\omega}|=\langle\phi^{\ast}|/\sqrt{2(\kappa^{2}-1)}$, and $|\omega\rangle=|\phi\rangle/\sqrt{2(\kappa^{2}-1)}$ to establish the PCR of Eq.~(\ref{SEq-3-1}).

\begin{table}[htpb]
	\centering
	\caption{Same as Table~\ref{tab1} except for $\epsilon=i$ and $\kappa\neq\pm1$.}
	\setlength{\tabcolsep}{6.07mm}{
		\begin{tabular}{c c c c c c c}\hline\hline
			& $|A\rangle$ & $|B\rangle$ & $|C\rangle$ & $|D\rangle$ & Left partner & Biorthogonal norm \\ \hline
			$|\psi_{1}\rangle$ & $-2i(\kappa^{2}-1)$ & 0 & $2(\kappa^{2}-1)$ & 0 & $\langle\psi_{3}^{\ast}|$ & $\sqrt{2(\kappa^{2}-1)}$ \\
			$|\psi_{2}\rangle$ & 0 & $i-\kappa$ & 0 & $i+\kappa$ & $\langle\psi_{2}^{\ast}|$ & $\sqrt{2(\kappa^{2}-1)}$ \\ 
			$|\psi_{3}\rangle$ & 0 & 0 & 1 & 0 & $\backslash$ & $\backslash$ \\ 
			$|\varphi_{3}\rangle$ & $i$ & 0 & 0 & 0 & $\langle\psi_{1}^{\ast}|$ & $\sqrt{2(\kappa^{2}-1)}$ \\ 
			$|\phi\rangle$ & 0 & $\kappa+i$ & 0 & $\kappa-i$ & $\langle\phi^{\ast}|$ & $\sqrt{2(\kappa^{2}-1)}$ \\ \hline\hline
		\end{tabular}
	}
	\label{tab3}
\end{table}

\begin{table}[htpb]
	\centering
	\caption{Same as Table~\ref{tab1} except for $\epsilon=-i$ and $\kappa\neq\pm1$.}
	\setlength{\tabcolsep}{6.07mm}{
		\begin{tabular}{c c c c c c c}\hline\hline
			& $|A\rangle$ & $|B\rangle$ & $|C\rangle$ & $|D\rangle$ & Left partner & Biorthogonal norm \\ \hline
			$|\psi_{1}\rangle$ & $-2i(\kappa^{2}-1)$ & 0 & $2(1-\kappa^{2})$ & 0 & $\langle\psi_{3}^{\ast}|$ & $\sqrt{2(\kappa^{2}-1)}$ \\ 
			$|\psi_{2}\rangle$ & 0 & $i-\kappa$ & 0 & $i+\kappa$ & $\langle\psi_{2}^{\ast}|$ & $\sqrt{2(\kappa^{2}-1)}$ \\ 
			$|\psi_{3}\rangle$ & 0 & 0 & $-1$ & 0 & $\backslash$ & $\backslash$ \\ 
			$|\varphi_{3}\rangle$ & $i$ & 0 & 0 & 0 & $\langle\psi_{1}^{\ast}|$ & $\sqrt{2(\kappa^{2}-1)}$ \\ 
			$|\phi\rangle$ & 0 & $\kappa+i$ & 0 & $\kappa-i$ & $\langle\phi^{\ast}|$ & $\sqrt{2(\kappa^{2}-1)}$ \\ \hline\hline
		\end{tabular}
	}
	\label{tab4}
\end{table}

We now consider the case of $\epsilon=\pm i$ and $\kappa=\pm1$. For $\epsilon=i$ and $\kappa=1$, the unnormalized eigenstates $|\psi_{1}^{(1)}\rangle$ and $|\psi_{1}^{(2)}\rangle$ and generalized eigenstates $|\psi_{2}^{(1)}\rangle$ and $|\psi_{2}^{(2)}\rangle$ of the two degenerate 2nd-order EPs are given in Table~\ref{tab5}, with their matching left partners $\langle\psi_{2}^{(2)\ast}|$, $\langle\psi_{2}^{(1)\ast}|$, $\langle\psi_{1}^{(2)\ast}|$, and $\langle\psi_{1}^{(1)\ast}|$. We need to rewrite $|\psi_{2}^{(1)}\rangle$ and $|\psi_{2}^{(2)}\rangle$ as $|\varphi_{2}^{(1)}\rangle$ and $|\varphi_{2}^{(2)}\rangle$, respectively, to hold the biorthogonality, as shown in Table~\ref{tab5}. By further redefining $\langle\tilde{\nu}_{2}^{(2)}|=\langle\psi_{2}^{(2)\ast}|/\sqrt{2(1-i)}$, $|\nu_{1}^{(1)}\rangle=|\psi_{1}^{(1)}\rangle/\sqrt{2(1-i)}$,  $\langle\tilde{\nu}_{1}^{(2)}|=\langle\psi_{1}^{(2)\ast}|/\sqrt{2(1-i)}$, $|\nu_{2}^{(1)}\rangle=|\varphi_{2}^{(1)}\rangle/\sqrt{2(1-i)}$,  $\langle\tilde{\nu}_{2}^{(1)}|=\langle\psi_{2}^{(1)\ast}|/\sqrt{2(1-i)}$, $|\nu_{1}^{(2)}\rangle=|\psi_{1}^{(2)}\rangle/\sqrt{2(1-i)}$,  $\langle\tilde{\nu}_{1}^{(1)}|=\langle\psi_{1}^{(1)\ast}|/\sqrt{2(1-i)}$, and $|\nu_{2}^{(2)}\rangle=|\varphi_{2}^{(2)}\rangle/\sqrt{2(1-i)}$,   the PCR can be established as
\begin{equation}\label{SEq-3-2}
\begin{split}
|\nu_{1}^{(1)}\rangle\langle\tilde{\nu}_{2}^{(2)}|+|\nu_{2}^{(1)}\rangle\langle\tilde{\nu}_{1}^{(2)}|+|\nu_{1}^{(2)}\rangle\langle\tilde{\nu}_{2}^{(1)}|+|\nu_{2}^{(2)}\rangle\langle\tilde{\nu}_{1}^{(1)}|=\hat{\mathbb{I}}.
\end{split}
\end{equation}

\begin{table}[htpb]
	\centering
	\caption{Distributions of eigenstates $|\psi_{1}^{(1)}\rangle$ and $|\psi_{1}^{(2)}\rangle$, generalized eigenstates $|\psi_{2}^{(1)}\rangle$ and $|\psi_{2}^{(2)}\rangle$, and redefined biorthogonal generalized eigenstates $|\varphi_{2}^{(1)}\rangle$ and $|\varphi_{2}^{(2)}\rangle$ of system at the two degenerate 2nd-order EPs for $\epsilon=i$ and $\kappa=1$. Their matching left partners and the corresponding biorthogonal norms are also listed, respectively.}
	\setlength{\tabcolsep}{7.67mm}{
		\begin{tabular}{c c c c c c c}\hline\hline
			& $|A\rangle$ & $|B\rangle$ & $|C\rangle$ & $|D\rangle$ & Left partner & Biorthogonal norm \\ \hline
			$|\psi_{1}^{(1)}\rangle$ & 0 & $2i$ & 0 & 2 & $\langle\psi_{2}^{(2)\ast}|$ & $\sqrt{2(1-i)}$ \\ 
			$|\psi_{2}^{(1)}\rangle$ & 1 & 0 & 1 & 0 & $\backslash$ & $\backslash$ \\ 
			$|\varphi_{2}^{(1)}\rangle$ & $i$ & 0 & $-i$ & 0 & $\langle\psi_{1}^{(2)\ast}|$ & $\sqrt{2(1-i)}$ \\ 
			$|\psi_{1}^{(2)}\rangle$ & $-2i$ & 0 & 2 & 0 & $\langle\psi_{2}^{(1)\ast}|$ & $\sqrt{2(1-i)}$ \\ 
			$|\psi_{2}^{(2)}\rangle$ & 0 & $-1$ & 0 & 1 & $\backslash$ & $\backslash$ \\ 
			$|\varphi_{2}^{(2)}\rangle$ & 0 & $-i$ & 0 & $-i$ & $\langle\psi_{1}^{(1)\ast}|$ & $\sqrt{2(1-i)}$ \\ 
			\hline\hline
		\end{tabular}
	}
	\label{tab5}
\end{table}

In Table~\ref{tab6}, we also give $|\psi_{1}^{(1)}\rangle$, $|\psi_{2}^{(1)}\rangle$, $|\psi_{1}^{(2)}\rangle$, and $|\psi_{2}^{(2)}\rangle$ for $\epsilon=i$ and $\kappa=-1$, with their matching left partners $\langle\psi_{2}^{(2)\ast}|$, $\langle\psi_{1}^{(2)\ast}|$, $\langle\psi_{2}^{(1)\ast}|$, and $\langle\psi_{1}^{(1)\ast}|$. We still need to rewrite $|\psi_{2}^{(1)}\rangle$ and $|\psi_{2}^{(2)}\rangle$ as $|\varphi_{2}^{(1)}\rangle$ and $|\varphi_{2}^{(2)}\rangle$, and to further redefine $\langle\tilde{\nu}_{2}^{(2)}|=\langle\psi_{2}^{(2)\ast}|/\sqrt{2(-1-i)}$, $|\nu_{1}^{(1)}\rangle=|\psi_{1}^{(1)}\rangle/\sqrt{2(-1-i)}$, $\langle\tilde{\nu}_{1}^{(2)}|=\langle\psi_{1}^{(2)\ast}|/\sqrt{2(-1-i)}$, $|\nu_{2}^{(1)}\rangle=|\varphi_{2}^{(1)}\rangle/\sqrt{2(-1-i)}$, $\langle\tilde{\nu}_{2}^{(1)}|=\langle\psi_{2}^{(1)\ast}|/\sqrt{2(-1-i)}$, $|\nu_{1}^{(2)}\rangle=|\psi_{1}^{(2)}\rangle/\sqrt{2(-1-i)}$,  $\langle\tilde{\nu}_{1}^{(1)}|=\langle\psi_{1}^{(1)\ast}|/\sqrt{2(-1-i)}$, and $|\nu_{2}^{(2)}\rangle=|\varphi_{2}^{(2)}\rangle/\sqrt{2(-1-i)}$ to establish the PCR of Eq.~(\ref{SEq-3-2}).

\begin{table}[htpb]
	\centering
	\caption{Same as Table~\ref{tab5} except for $\kappa=-1$.}
	\setlength{\tabcolsep}{7.65mm}{
		\begin{tabular}{c c c c c c c}\hline\hline
			& $|A\rangle$ & $|B\rangle$ & $|C\rangle$ & $|D\rangle$ & Left partner & Biorthogonal norm \\ \hline
			$|\psi_{1}^{(1)}\rangle$ & 0 & $-2i$ & 0 & 2 & $\langle\psi_{2}^{(2)\ast}|$ & $\sqrt{2(-1-i)}$ \\ 
			$|\psi_{2}^{(1)}\rangle$ & 1 & 0 & $-1$ & 0 & $\backslash$ & $\backslash$ \\ 
			$|\varphi_{2}^{(1)}\rangle$ & $-i$ & 0 & $-i$ & 0 & $\langle\psi_{1}^{(2)\ast}|$ & $\sqrt{2(-1-i)}$ \\ 
			$|\psi_{1}^{(2)}\rangle$ & $-2i$ & 0 & 2 & 0 & $\langle\psi_{2}^{(1)\ast}|$ & $\sqrt{2(-1-i)}$ \\ 
			$|\psi_{2}^{(2)}\rangle$ & 0 & 1 & 0 & $-1$ & $\backslash$ & $\backslash$ \\ 
			$|\varphi_{2}^{(2)}\rangle$ & 0 & $-i$ & 0 & $-i$ & $\langle\psi_{1}^{(1)\ast}|$ & $\sqrt{2(-1-i)}$ \\ 
			\hline\hline
		\end{tabular}
	}
	\label{tab6}
\end{table}

Similarly, for $\epsilon=-i$ and $\kappa=\pm1$, the corresponding $|\psi_{1}^{(1)}\rangle$, $|\psi_{2}^{(1)}\rangle$, $|\psi_{1}^{(2)}\rangle$, and $|\psi_{2}^{(2)}\rangle$ are also given in Tables~\ref{tab7} and~\ref{tab8}, respectively, with their matching left partners $\langle\psi_{2}^{(2)\ast}|$, $\langle\psi_{1}^{(2)\ast}|$, $\langle\psi_{2}^{(1)\ast}|$, and $\langle\psi_{1}^{(1)\ast}|$. By rewriting $|\psi_{2}^{(1)}\rangle$ and $|\psi_{2}^{(2)}\rangle$ as $|\varphi_{2}^{(1)}\rangle$ and $|\varphi_{2}^{(2)}\rangle$, and further redefining $\langle\tilde{\nu}_{2}^{(2)}|=\langle\psi_{2}^{(2)\ast}|/\sqrt{2(i-1)}$ and $\langle\psi_{2}^{(2)\ast}|/\sqrt{2(1+i)}$, $|\nu_{1}^{(1)}\rangle=|\psi_{1}^{(1)}\rangle/\sqrt{2(i-1)}$ and $|\psi_{1}^{(1)}\rangle/\sqrt{2(1+i)}$, $\langle\tilde{\nu}_{1}^{(2)}|=\langle\psi_{1}^{(2)\ast}|/\sqrt{2(i-1)}$ and $\langle\psi_{1}^{(2)\ast}|/\sqrt{2(1+i)}$, $|\nu_{2}^{(1)}\rangle=|\varphi_{2}^{(1)}\rangle/\sqrt{2(i-1)}$ and $|\varphi_{2}^{(1)}\rangle/\sqrt{2(1+i)}$, $\langle\tilde{\nu}_{2}^{(1)}|=\langle\psi_{2}^{(1)\ast}|/\sqrt{2(i-1)}$ and $\langle\psi_{2}^{(1)\ast}|/\sqrt{2(1+i)}$, $|\nu_{1}^{(2)}\rangle=|\psi_{1}^{(2)}\rangle/\sqrt{2(i-1)}$ and $|\psi_{1}^{(2)}\rangle/\sqrt{2(1+i)}$, $\langle\tilde{\nu}_{1}^{(1)}|=\langle\psi_{1}^{(1)\ast}|/\sqrt{2(i-1)}$ and $\langle\psi_{1}^{(1)\ast}|/\sqrt{2(1+i)}$, and $|\nu_{2}^{(2)}\rangle=|\varphi_{2}^{(2)}\rangle/\sqrt{2(i-1)}$ and $|\varphi_{2}^{(2)}\rangle/\sqrt{2(1+i)}$, we can still establish the PCR of Eq.~(\ref{SEq-3-2}).

\begin{table}[htpb]
	\centering
	\caption{Same as Table~\ref{tab5} except for $\epsilon=-i$.}
	\setlength{\tabcolsep}{7.67mm}{
		\begin{tabular}{c c c c c c c}\hline\hline
			& $|A\rangle$ & $|B\rangle$ & $|C\rangle$ & $|D\rangle$ & Left partner & Biorthogonal norm \\ \hline
			$|\psi_{1}^{(1)}\rangle$ & 0 & $2i$ & 0 & 2 & $\langle\psi_{2}^{(2)\ast}|$ & $\sqrt{2(i-1)}$ \\ 
			$|\psi_{2}^{(1)}\rangle$ & 1 & 0 & $-1$ & 0 & $\backslash$ & $\backslash$ \\ 
			$|\varphi_{2}^{(1)}\rangle$ & $i$ & 0 & $i$ & 0 & $\langle\psi_{1}^{(2)\ast}|$ & $\sqrt{2(i-1)}$ \\ 
			$|\psi_{1}^{(2)}\rangle$ & $2i$ & 0 & 2 & 0 & $\langle\psi_{2}^{(1)\ast}|$ & $\sqrt{2(i-1)}$ \\ 
			$|\psi_{2}^{(2)}\rangle$ & 0 & 1 & 0 & $-1$ & $\backslash$ & $\backslash$ \\ 
			$|\varphi_{2}^{(2)}\rangle$ & 0 & $i$ & 0 & $i$ & $\langle\psi_{1}^{(1)\ast}|$ & $\sqrt{2(i-1)}$ \\ 
			\hline\hline
		\end{tabular}
	}
	\label{tab7}
\end{table}

\begin{table}[htpb]
	\centering
	\caption{Same as Table~\ref{tab5} except for $\epsilon=-i$ and $\kappa=-1$.}
	\setlength{\tabcolsep}{7.67mm}{
		\begin{tabular}{c c c c c c c}\hline\hline
			& $|A\rangle$ & $|B\rangle$ & $|C\rangle$ & $|D\rangle$ & Left partner & Biorthogonal norm \\ \hline
			$|\psi_{1}^{(1)}\rangle$ & 0 & $-2i$ & 0 & 2 & $\langle\psi_{2}^{(2)\ast}|$ & $\sqrt{2(1+i)}$ \\ 
			$|\psi_{2}^{(1)}\rangle$ & 1 & 0 & 1 & 0 & $\backslash$ & $\backslash$ \\ 
			$|\varphi_{2}^{(1)}\rangle$ & $-i$ & 0 & $i$ & 0 & $\langle\psi_{1}^{(2)\ast}|$ & $\sqrt{2(1+i)}$ \\ 
			$|\psi_{1}^{(2)}\rangle$ & $2i$ & 0 & 2 & 0 & $\langle\psi_{2}^{(1)\ast}|$ & $\sqrt{2(1+i)}$ \\ 
			$|\psi_{2}^{(2)}\rangle$ & 0 & $-1$ & 0 & 1 & $\backslash$ & $\backslash$ \\ 
			$|\varphi_{2}^{(2)}\rangle$ & 0 & $i$ & 0 & $i$ & $\langle\psi_{1}^{(1)\ast}|$ & $\sqrt{2(1+i)}$ \\ 
			\hline\hline
		\end{tabular}
	}
	\label{tab8}
\end{table}

\section{Identification of exceptional points in the two non-Hermitian models via Petermann factor}

\begin{figure}\centering
	\includegraphics[width=0.9\linewidth]{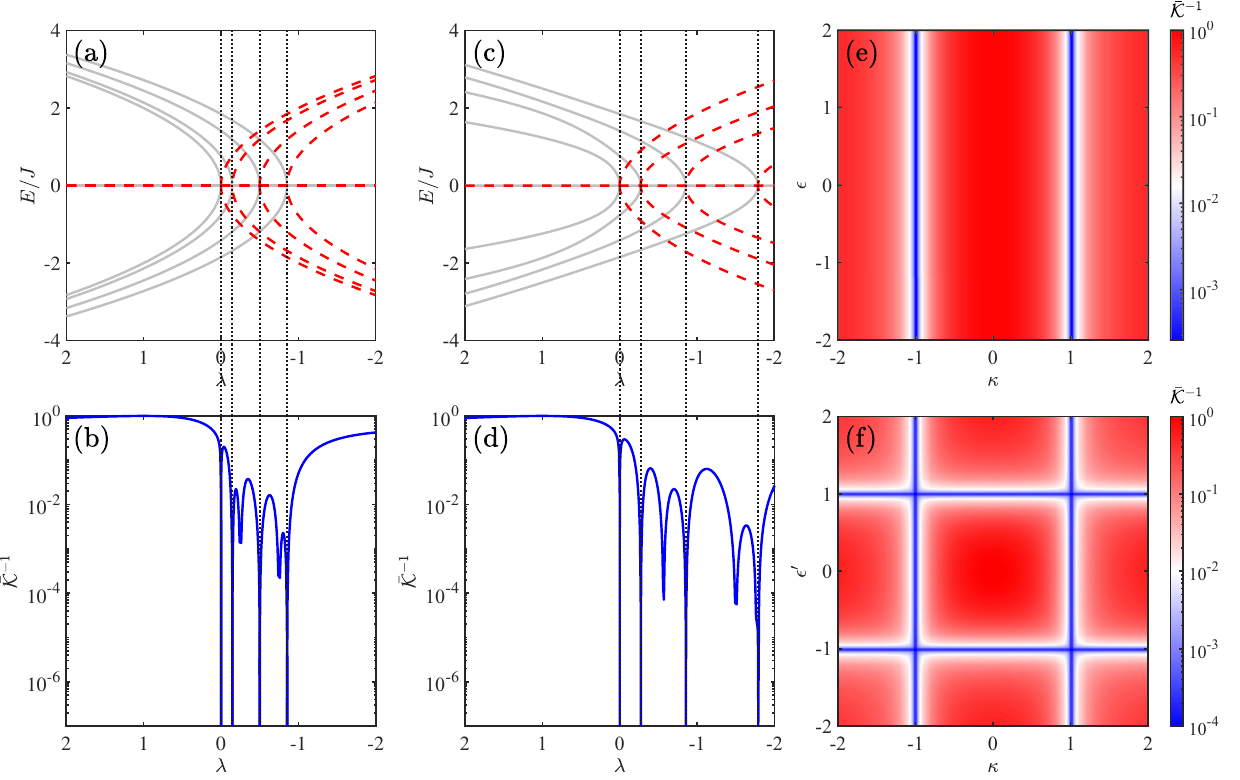}
	\caption{(a) Spectrum of NH stub ribbon with $N=4$ as a function of $\lambda$ when $\lbrace J_{n}^{u}=2J\rbrace$, where the gray-solid and red-dashed lines delineate the real and imaginary parts of spectrum, respectively. (b) Inverse average Petermann factor $\bar{\mathcal{K}}^{-1}$ versus $\lambda$ for the NH stub ribbon in (a). (c) and (d) Same as (a) and (b) except for $\lbrace J_{n}^{u}=\sqrt{n}J\rbrace$. The divergence of $\bar{\mathcal{K}}$ identifies the emergence of EPs, as guided by the black-dotted lines. (e) Inverse average Petermann factor $\bar{\mathcal{K}}^{-1}$ versus $\kappa$ and real $\epsilon$ for the NH diamond ring. (f) Same as (e) except for imaginary $\epsilon=i\epsilon^{\prime}$. $\bar{\mathcal{K}}$ diverges at $\kappa=\pm1$ or $\epsilon=\pm i$, where EPs are always present.}\label{fig4}
\end{figure}

A powerful tool to identify the emergence of EPs is the celebrated Petermann factor~\cite{1070064,PhysRevA.39.1253,PhysRevA.61.023810,Berry01012003,PhysRevA.78.015805,PhysRevA.82.010103,PhysRevResearch.2.032057}, which measures the nonorthogonality of the eigenstates of a NH system and is responsible for the enhancing intrinsic linewidth of lasers, defined as 
\begin{equation}\label{SEq-4-1}
\mathcal{K}_{m}=\frac{\langle L_{m}|L_{m}\rangle\langle R_{m}|R_{m}\rangle}{\left|\langle L_{m}|R_{m}\rangle\right|^{2}},
\end{equation}	
for the $m$th right eigenstate $|R_{m}\rangle$ and its left partner $|L_{m}\rangle$. We further consider the average Petermann factor
\begin{equation}\label{SEq-4-2}
\bar{\mathcal{K}}=\frac{1}{M}\sum_{m=1}^{M}\mathcal{K}_{m},
\end{equation}
with $M$ the dimension of the Hilbert space. It will deviate from one due to the violation of the standard orthonormalization and become divergent in the immediate vicinity of EPs.    

Figures~\ref{fig4}(a) and~\ref{fig4}(c) show the spectra of the NH stub ribbon with $N=4$ in a larger range of $\lambda$ when $\lbrace J_{n}^{u}=2J\rbrace$ and $\lbrace J_{n}^{u}=\sqrt{n}J\rbrace$, where the real and imaginary parts of each spectrum are depicted by the gray-solid and red-dashed lines, respectively. One can observe that the system admits a real spectrum for $\lambda>0$, since the model is always similar to a Hermitian stub ribbon, $\hat{S}^{-1}\hat{H}\hat{S}=\hat{H}_{h}$, with the transformation matrix $\hat{S}=\sqrt{\hat{\eta}}$, as given in the main text. Additionally, as mentioned in the main text, there exists a 2nd-order EP at $\lambda=0$, where a couple of real eigenenergies merge into the zero energy and their eigenstates coalesce, leading to a hybridization of one 3rd-order DP and one 2nd-order EP. However, the imaginary eigenenergies will occur once $\lambda<0$, but the pseudo-Hermiticity, $\hat{\eta}^{-1}\hat{H}\hat{\eta}=\hat{H}^{\dagger}$, is still respected, so that they will appear in complex conjugate pairs. As $\lambda$ gradually decreases, the remaining three couples of real eigenenergies merge into the zero energy successively at different $\lambda$, forming a 3rd-order EP together with one of the three original degenerate zero eigenenergies and thus yielding a hybridization of one 2nd-order DP and one 3rd-order EP. 

We further show in Figs.~\ref{fig4}(b) and~\ref{fig4}(d) the corresponding inverse average Petermann factors $\bar{\mathcal{K}}^{-1}$ as a function of $\lambda$, respectively, for the two NH stub ribbons in Figs.~\ref{fig4}(a) and~\ref{fig4}(c). It is clear that $\bar{\mathcal{K}}$ diverges at these specific values of $\lambda$, which corroborates the presence of these EPs. Otherwise, $\bar{\mathcal{K}}$ always remains finite. 

Similarly, for the NH diamond ring, the inverse average Petermann factors $\bar{\mathcal{K}}^{-1}$ versus $\kappa$ and real $\epsilon$ or imaginary $\epsilon=i\epsilon^{\prime}$ are shown in Figs.~\ref{fig4}(e) and~\ref{fig4}(f), respectively, where the divergence of $\bar{\mathcal{K}}$ at $\kappa=\pm1$ or $\epsilon=\pm i$ also reflects the existence of EPs.

\section{Experimental implementation of nonreciprocal hopping and complex coupling}

\begin{figure}\centering
	\includegraphics[width=0.85\linewidth]{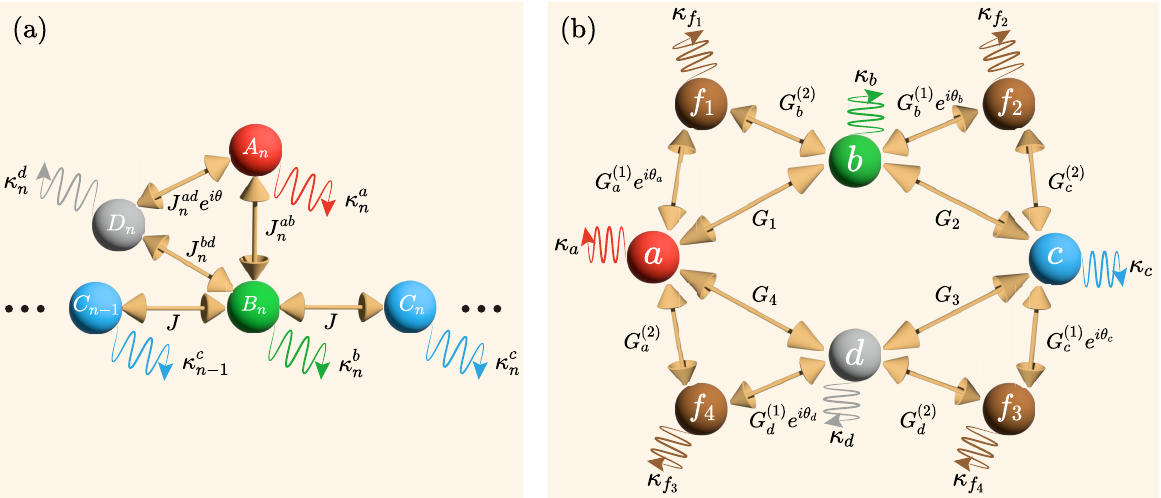}
	\caption{(a) Sketch of the experimental implementation of the NH stub ribbon with nonreciprocal hopping in a coupled-cavity array, where the cavities $A_{n}$ and $B_{n}$ are coupled indirectly via the auxiliary cavity $D_{n}$ with high dissipation and the asymmetric coupling between them can be effectively realized after eliminating the mode of the cavity $D_{n}$ adiabatically. (b) Same as (a) but for the NH diamond ring with complex coupling, where there are four highly dissipative auxiliary modes $f_{1}$, $f_{2}$, $f_{3}$, and $f_{4}$. In each panel, the corresponding coupling strength and dissipation rate have been indicated, respectively.}\label{fig5}
\end{figure}

Experimentally, the NH stub ribbon with the nonreciprocity emerging in the nearest-neighbor hopping can be mapped to a coupled-mode quantum system~\cite{PhysRevApplied.13.044070,Du2020}, such as a coupled-cavity array with the configuration shown in Fig.~\ref{fig5}(a), where the cavities $A_{n}$, $B_{n}$, and $C_{n}$ possess single modes $a_{n}$, $b_{n}$, and $c_{n}$ of the identical frequency $\varepsilon$ and the dissipation rates $\kappa_{n}^{a}$, $\kappa_{n}^{b}$, and $\kappa_{n}^{c}$, and play the roles of the sublattices $A$, $B$, and $C$ within the $n$th unit cell, respectively. In addition, there is another auxiliary cavity $D_{n}$ with single mode $d_{n}$, whose frequency and dissipation rate are $\varepsilon$ and $\kappa_{n}^{d}$, respectively, inside the same unit cell $n$. The cavities $A_{n}$ and $B_{n}$ are coupled directly with the coupling strength $J_{n}^{ab}$ and indirectly via the auxiliary cavity $D_{n}$, where the coupling strength between the cavities $A_{n}$ $(B_{n})$ and $D_{n}$ is $J_{n}^{ad}e^{i\theta}$ $(J_{n}^{bd})$. The cavities $B_{n}$ $(B_{n+1})$ and $C_{n}$ are coupled to each other with the coupling strength $J$. The Hamiltonian of the coupled-cavity array thus reads 
\begin{equation}\label{SEq-5-1}
\begin{split}
\hat{H}=&\sum_{n=1}^{N}\varepsilon\left(\hat{a}_{n}^{\dagger}\hat{a}_{n}+\hat{b}_{n}^{\dagger}\hat{b}_{n}+\hat{c}_{n}^{\dagger}\hat{c}_{n}+\hat{d}_{n}^{\dagger}\hat{d}_{n}\right)-\varepsilon\hat{c}_{N}^{\dagger}\hat{c}_{N}\\
&+\sum_{n=1}^{N}\left(J_{n}^{ab}\hat{a}_{n}^{\dagger}\hat{b}_{n}+J_{n}^{ad}e^{i\theta}\hat{a}_{n}^{\dagger}\hat{d}_{n}+J_{n}^{bd}\hat{b}_{n}^{\dagger}\hat{d}_{n}+\mathrm{H.c.}\right)\\
&+\sum_{n=1}^{N-1}J\left(\hat{c}_{n}^{\dagger}\hat{b}_{n}+\hat{c}_{n}^{\dagger}\hat{b}_{n+1}+\mathrm{H.c.}\right),
\end{split}
\end{equation}
where $\hat{a}_{n}^{\dagger}$, $\hat{b}_{n}^{\dagger}$, $\hat{c}_{n}^{\dagger}$, and $\hat{d}_{n}^{\dagger}$ denote the creation operators of the corresponding cavity modes, respectively, and the first line of Eq.~(\ref{SEq-5-1}) represents their free-energy terms. In the interaction picture with respect to the frequency $\varepsilon$, the above Hamiltonian is transformed into
\begin{equation}\label{SEq-5-2}
\begin{split}
\hat{H}^{'}=\sum_{n=1}^{N}\left(J_{n}^{ab}\hat{a}_{n}^{\dagger}\hat{b}_{n}+J_{n}^{ad}e^{i\theta}\hat{a}_{n}^{\dagger}\hat{d}_{n}+J_{n}^{bd}\hat{b}_{n}^{\dagger}\hat{d}_{n}+\mathrm{H.c.}\right)+\sum_{n=1}^{N-1}J\left(\hat{c}_{n}^{\dagger}\hat{b}_{n}+\hat{c}_{n}^{\dagger}\hat{b}_{n+1}+\mathrm{H.c.}\right).
\end{split}
\end{equation}

In the regime of dissipation, the motion equation of each mode can be described as
\begin{equation}\label{SEq-5-3}
\begin{split}
\dot{\hat{a}}_{n}=&-iJ_{n}^{ab}\hat{b}_{n}-iJ_{n}^{ad}e^{i\theta}\hat{d}_{n}-\frac{\kappa_{n}^{a}}{2}\hat{a}_{n},\\
\dot{\hat{b}}_{n}=&-iJ_{n}^{ab}\hat{a}_{n}-iJ\Big(\hat{c}_{n-1}+\hat{c}_{n}\Big)-iJ_{n}^{bd}\hat{d}_{n}-\frac{\kappa_{n}^{b}}{2}\hat{b}_{n},\\
\dot{\hat{c}}_{n}=&-iJ\Big(\hat{b}_{n}+\hat{b}_{n+1}\Big)-\frac{\kappa_{n}^{c}}{2}\hat{c}_{n},\\
\dot{\hat{d}}_{n}=&-iJ_{n}^{ad}e^{-i\theta}\hat{a}_{n}-iJ_{n}^{bd}\hat{b}_{n}-\frac{\kappa_{n}^{d}}{2}\hat{d}_{n}.
\end{split}
\end{equation}
If $\kappa_{n}^{d}$ is sufficiently large, the mode $d_{n}$ can be treated as a dissipative bath and be removed adiabatically by taking $\dot{\hat{d}}_{n}=0$, further yielding
\begin{equation}\label{SEq-5-4}
\hat{d}_{n}=-i\frac{2}{\kappa_{n}^{d}}J_{n}^{ad}e^{-i\theta}\hat{a}_{n}-i\frac{2}{\kappa_{n}^{d}}J_{n}^{bd}\hat{b}_{n},
\end{equation}
and the motion equations for the modes $a_{n}$, $b_{n}$, and $c_{n}$ thereby become
\begin{equation}\label{Eq-5-5}
\begin{split}
\dot{\hat{a}}_{n}=&-i\left(J_{n}^{ab}-i\tilde{J}_{n}^{ab}e^{i\theta}\right)\hat{b}_{n}-\frac{1}{2}\Big(\kappa_{n}^{a}+\tilde{\kappa}_{n}^{a}\Big)\hat{a}_{n},\\
\dot{\hat{b}}_{n}=&-i\left(J_{n}^{ab}-i\tilde{J}_{n}^{ab}e^{-i\theta}\right)\hat{a}_{n}-iJ\Big(\hat{c}_{n-1}+\hat{c}_{n}\Big)-\frac{1}{2}\Big(\kappa_{n}^{b}+\tilde{\kappa}_{n}^{b}\Big)\hat{b}_{n},\\
\dot{\hat{c}}_{n}=&-iJ\Big(\hat{b}_{n}+\hat{b}_{n+1}\Big)-\frac{\kappa_{n}^{c}}{2}\hat{c}_{n},
\end{split}
\end{equation}
where $\tilde{J}_{n}^{ab}=2J_{n}^{ad}J_{n}^{bd}/\kappa_{n}^{d}$, $\tilde{\kappa}_{n}^{a}=4(J_{n}^{ad})^{2}/\kappa_{n}^{d}$, and $\tilde{\kappa}_{n}^{b}=4(J_{n}^{bd})^{2}/\kappa_{n}^{d}$. Accordingly, we can obtain the final effective Hamiltonian
\begin{equation}\label{SEq-5-6}
\begin{split}
\hat{H}_{\mathrm{eff}}=\sum_{n=1}^{N}\left[\left(J_{n}^{ab}-i\tilde{J}_{n}^{ab}e^{i\theta}\right)\hat{a}_{n}^{\dagger}\hat{b}_{n}+\left(J_{n}^{ab}-i\tilde{J}_{n}^{ab}e^{-i\theta}\right)\hat{b}_{n}^{\dagger}\hat{a}_{n}\right]+\sum_{n=1}^{N-1}J\left(\hat{c}_{n}^{\dagger}\hat{b}_{n}+\hat{c}_{n}^{\dagger}\hat{b}_{n+1}+\mathrm{H.c.}\right).
\end{split}
\end{equation}
When $\theta=\pi/2$, equation~(\ref{SEq-5-6}) can be further simplified as
\begin{equation}\label{SEq-5-7}
\begin{split}
\hat{H}_{\mathrm{eff}}^{\prime}=\sum_{n=1}^{N}\left[\left(J_{n}^{ab}+\tilde{J}_{n}^{ab}\right)\hat{a}_{n}^{\dagger}\hat{b}_{n}+\left(J_{n}^{ab}-\tilde{J}_{n}^{ab}\right)\hat{b}_{n}^{\dagger}\hat{a}_{n}\right]+\sum_{n=1}^{N-1}J\left(\hat{c}_{n}^{\dagger}\hat{b}_{n}+\hat{c}_{n}^{\dagger}\hat{b}_{n+1}+\mathrm{H.c.}\right),
\end{split}
\end{equation}
leading to the asymmetric coupling between the cavities $A_{n}$ and $B_{n}$ assisted by the auxiliary cavity $D_{n}$ with high dissipation. If we set $J_{n}^{ab}+\tilde{J}_{n}^{ab}=J_{n}^{u}$ and $J_{n}^{ab}-\tilde{J}_{n}^{ab}=J_{n}^{d}$, the NH stub ribbon with the nonreciprocal hopping between the sublattices $A$ and $B$ in the same unit cell is implemented and the 2nd-order EP can be generated for $J_{n}^{ab}=\tilde{J}_{n}^{ab}$.

Similarly, we can also introduce four auxiliary modes $f_{1}$, $f_{2}$, $f_{3}$, and $f_{4}$ to implement the complex coupling in the NH diamond ring, as shown in Fig.~\ref{fig5}(b), where the coupling strength between each two modes and the dissipation rate of each mode have been marked, respectively, and all of the modes possess the identical frequency $\varepsilon$. In the interaction picture with respect to the frequency $\varepsilon$, the Hamiltonian of the system can be written as
\begin{equation}\label{SEq-5-8}
\begin{split}
\hat{H}=&G_{1}\hat{a}^{\dagger}\hat{b}+G_{2}\hat{b}^{\dagger}\hat{c}+G_{a}^{(1)}e^{i\theta_{a}}\hat{a}^{\dagger}\hat{f}_{1}+G_{b}^{(2)}\hat{b}^{\dagger}\hat{f}_{1}+G_{b}^{(1)}e^{i\theta_{b}}\hat{b}^{\dagger}\hat{f}_{2}+G_{c}^{(2)}\hat{c}^{\dagger}\hat{f}_{2}\\
&+G_{3}\hat{c}^{\dagger}\hat{d}+G_{4}\hat{d}^{\dagger}\hat{a}+G_{c}^{(1)}e^{i\theta_{c}}\hat{c}^{\dagger}\hat{f}_{3}+G_{d}^{(2)}\hat{d}^{\dagger}\hat{f}_{3}+G_{d}^{(1)}e^{i\theta_{d}}\hat{d}^{\dagger}\hat{f}_{4}+G_{a}^{(2)}\hat{a}^{\dagger}\hat{f}_{4}+\mathrm{H.c.},
\end{split}
\end{equation}
and in the regime of dissipation, the motion equations for all of the modes read
\begin{equation}\label{SEq-5-9}
\begin{split}
\dot{\hat{a}}=&-iG_{1}\hat{b}-iG_{4}\hat{d}-iG_{a}^{(1)}e^{i\theta_{a}}\hat{f}_{1}-iG_{a}^{(2)}\hat{f}_{4}-\frac{\kappa_{a}}{2}\hat{a},\\
\dot{\hat{b}}=&-iG_{1}\hat{a}-iG_{2}\hat{c}-iG_{b}^{(1)}e^{i\theta_{b}}\hat{f}_{2}-iG_{b}^{(2)}\hat{f}_{1}-\frac{\kappa_{b}}{2}\hat{b},\\
\dot{\hat{c}}=&-iG_{2}\hat{b}-iG_{3}\hat{d}-iG_{c}^{(1)}e^{i\theta_{c}}\hat{f}_{3}-iG_{c}^{(2)}\hat{f}_{2}-\frac{\kappa_{c}}{2}\hat{c},\\
\dot{\hat{d}}=&-iG_{3}\hat{c}-iG_{4}\hat{a}-iG_{d}^{(1)}e^{i\theta_{d}}\hat{f}_{4}-iG_{d}^{(2)}\hat{f}_{3}-\frac{\kappa_{d}}{2}\hat{d},\\
\dot{\hat{f}}_{1}=&-iG_{a}^{(1)}e^{-i\theta_{a}}\hat{a}-iG_{b}^{(2)}\hat{b}-\frac{\kappa_{f_{1}}}{2}\hat{f}_{1},\\
\dot{\hat{f}}_{2}=&-iG_{b}^{(1)}e^{-i\theta_{b}}\hat{b}-iG_{c}^{(2)}\hat{c}-\frac{\kappa_{f_{2}}}{2}\hat{f}_{2},\\
\dot{\hat{f}}_{3}=&-iG_{c}^{(1)}e^{-i\theta_{c}}\hat{c}-iG_{d}^{(2)}\hat{d}-\frac{\kappa_{f_{3}}}{2}\hat{f}_{3},\\
\dot{\hat{f}}_{4}=&-iG_{d}^{(1)}e^{-i\theta_{d}}\hat{d}-iG_{a}^{(2)}\hat{a}-\frac{\kappa_{f_{4}}}{2}\hat{f}_{4}.
\end{split}
\end{equation}
When $\kappa_{f_{1}}$, $\kappa_{f_{2}}$, $\kappa_{f_{3}}$, and $\kappa_{f_{4}}$ are sufficiently large, we also take $\dot{\hat{f}}_{1}=\dot{\hat{f}}_{2}=\dot{\hat{f}}_{3}=\dot{\hat{f}}_{4}=0$ and further obtain
\begin{equation}\label{SEq-5-10}
\begin{split}
\dot{\hat{a}}=&-i\left(G_{1}-i\tilde{G}_{1}e^{i\theta_{a}}\right)\hat{b}-i\left(G_{4}-i\tilde{G}_{4}e^{-i\theta_{d}}\right)\hat{d}-\frac{1}{2}\Big(\kappa_{a}+\tilde{\kappa}_{a}^{(1)}+\tilde{\kappa}_{a}^{(2)}\Big)\hat{a},\\
\dot{\hat{b}}=&-i\left(G_{1}-i\tilde{G}_{1}e^{-i\theta_{a}}\right)\hat{a}-i\left(G_{2}-i\tilde{G}_{2}e^{i\theta_{b}}\right)\hat{c}-\frac{1}{2}\Big(\kappa_{b}+\tilde{\kappa}_{b}^{(1)}+\tilde{\kappa}_{b}^{(2)}\Big)\hat{b},\\
\dot{\hat{c}}=&-i\left(G_{2}-i\tilde{G}_{2}e^{-i\theta_{b}}\right)\hat{b}-i\left(G_{3}-i\tilde{G}_{3}e^{i\theta_{c}}\right)\hat{d}-\frac{1}{2}\Big(\kappa_{c}+\tilde{\kappa}_{c}^{(1)}+\tilde{\kappa}_{c}^{(2)}\Big)\hat{c},\\
\dot{\hat{d}}=&-i\left(G_{3}-i\tilde{G}_{3}e^{-i\theta_{c}}\right)\hat{c}-i\left(G_{4}-i\tilde{G}_{4}e^{i\theta_{d}}\right)\hat{a}-\frac{1}{2}\Big(\kappa_{d}+\tilde{\kappa}_{d}^{(1)}+\tilde{\kappa}_{d}^{(2)}\Big)\hat{d},
\end{split}
\end{equation}
with $\tilde{G}_{1}=2G_{a}^{(1)}G_{b}^{(2)}/\kappa_{f_{1}}$, $\tilde{G}_{2}=2G_{b}^{(1)}G_{c}^{(2)}/\kappa_{f_{2}}$, $\tilde{G}_{3}=2G_{c}^{(1)}G_{d}^{(2)}/\kappa_{f_{3}}$, $\tilde{G}_{4}=2G_{d}^{(1)}G_{a}^{(2)}/\kappa_{f_{4}}$, $\tilde{\kappa}_{a}^{(1)}=4(G_{a}^{(1)})^{2}/\kappa_{f_{1}}$, $\tilde{\kappa}_{a}^{(2)}=4(G_{a}^{(2)})^{2}/\kappa_{f_{4}}$, $\tilde{\kappa}_{b}^{(1)}=4(G_{b}^{(1)})^{2}/\kappa_{f_{2}}$, $\tilde{\kappa}_{b}^{(2)}=4(G_{b}^{(2)})^{2}/\kappa_{f_{1}}$, $\tilde{\kappa}_{c}^{(1)}=4(G_{c}^{(1)})^{2}/\kappa_{f_{3}}$, $\tilde{\kappa}_{c}^{(2)}=4(G_{c}^{(2)})^{2}/\kappa_{f_{2}}$, $\tilde{\kappa}_{d}^{(1)}=4(G_{d}^{(1)})^{2}/\kappa_{f_{4}}$, and $\tilde{\kappa}_{d}^{(2)}=4(G_{d}^{(2)})^{2}/\kappa_{f_{3}}$. Accordingly, the resulting effective Hamiltonian can be described as 
\begin{equation}\label{SEq-5-11}
\begin{split}
\hat{H}_{\mathrm{eff}}=&\left(G_{1}-i\tilde{G}_{1}e^{i\theta_{a}}\right)\hat{a}^{\dagger}\hat{b}+\left(G_{1}-i\tilde{G}_{1}e^{-i\theta_{a}}\right)\hat{b}^{\dagger}\hat{a}+\left(G_{2}-i\tilde{G}_{2}e^{i\theta_{b}}\right)\hat{b}^{\dagger}\hat{c}+\left(G_{2}-i\tilde{G}_{2}e^{-i\theta_{b}}\right)\hat{c}^{\dagger}\hat{b}\\
&+\left(G_{3}-i\tilde{G}_{3}e^{i\theta_{c}}\right)\hat{c}^{\dagger}\hat{d}+\left(G_{3}-i\tilde{G}_{3}e^{-i\theta_{c}}\right)\hat{d}^{\dagger}\hat{c}+\left(G_{4}-i\tilde{G}_{4}e^{i\theta_{d}}\right)\hat{d}^{\dagger}\hat{a}+\left(G_{4}-i\tilde{G}_{4}e^{-i\theta_{d}}\right)\hat{a}^{\dagger}\hat{d}.
\end{split}
\end{equation}
If we assign $\theta_{a}=\theta_{b}=\pi$, $\theta_{c}=\theta_{d}=0$, $G_{1}=G_{4}=J$, $\tilde{G}_{1}=\tilde{G}_{4}=\kappa$, $G_{2}=G_{3}=\epsilon J$, and $\tilde{G}_{2}=\tilde{G}_{3}=\epsilon\kappa$, equation~(\ref{SEq-5-11}) becomes
\begin{equation}\label{SEq-5-12}
\begin{split}
\hat{H}_{\mathrm{eff}}^{\prime}=&\left(J+i\kappa\right)\left(\hat{a}^{\dagger}\hat{b}+\hat{b}^{\dagger}\hat{a}\right)+\epsilon\left(J+i\kappa\right)\left(\hat{b}^{\dagger}\hat{c}+\hat{c}^{\dagger}\hat{b}\right)\\
&+\epsilon\left(J-i\kappa\right)\left(\hat{c}^{\dagger}\hat{d}+\hat{d}^{\dagger}\hat{c}\right)+\left(J-i\kappa\right)\left(\hat{d}^{\dagger}\hat{a}+\hat{a}^{\dagger}\hat{d}\right),
\end{split}
\end{equation}
which is mapped to the NH diamond ring with complex coupling for real $\epsilon$. For imaginary $\epsilon=i\epsilon^{\prime}$, we need to reset $\tilde{G}_{2}=-\tilde{G}_{3}=\epsilon^{\prime}J$ and $-G_{2}=G_{3}=\epsilon^{\prime}\kappa$.


\end{document}